\documentclass{IEEEtran}
\usepackage{amsmath,amsfonts}
\usepackage{algorithm}
\usepackage{algpseudocode}
\usepackage{array}
\usepackage{hyperref}
\usepackage{multirow} 
\usepackage[font=small]{caption}
\usepackage{subcaption}
\usepackage{textcomp}
\usepackage{stfloats}
\usepackage{booktabs}
\usepackage{amssymb} 
\usepackage{siunitx}
\usepackage{url}
\usepackage{verbatim}
\usepackage{enumitem} 
\usepackage{graphicx}
\usepackage[monochrome]{xcolor} 
\usepackage{cite}
\usepackage{kotex} 
\usepackage[normalem]{ulem}
\usepackage{bm} 
\newcommand{\soutMath}[1]{\ifmmode\text{\sout{\ensuremath{#1}}}\else\sout{#1}\fi}

\usepackage{amsthm,thmtools} 
\newtheorem{remark}{Remark}
\newtheorem{lemma}{Lemma}
\newtheorem{corollary}{Corollary}

\newtheorem{assumption}{Assumption}
\newtheorem{theorem}{Theorem}
\newtheorem{claim}{Claim}
\newtheorem{problem}{Problem}
\newtheorem{definition}{Definition}

\newcommand\red[1]{\textcolor{red}{#1}}
\newcommand\blue[1]{\textcolor{blue}{#1}}
\DeclareMathOperator*{\argmax}{argmax}
\DeclareMathOperator*{\argmin}{argmin}

\usepackage{textcomp}
\def\BibTeX{{\rm B\kern-.05em{\sc i\kern-.025em b}\kern-.08em
    T\kern-.1667em\lower.7ex\hbox{E}\kern-.125emX}}
    
\begin{document}

\title{Switching control of underactuated multi-channel systems with input constraints for cooperative manipulation}


\author{Dongjae Lee, Dimos V. Dimarogonas, and H. Jin Kim
\thanks{D. Lee, H. J. Kim are with the Department of Aerospace Engineering and the Automation and Systems Research Institute (ASRI), Seoul National University (SNU), Seoul 08826, South Korea {\tt\small \{ehdwo713, hjinkim\}@snu.ac.kr}}%
\thanks{D. V. Dimarogonas is with the Division of Decision and Control Systems, KTH Royal Institute of Technology, Stockholm, Sweden {\tt\small dimos@kth.se}}
}



\maketitle

\begin{abstract}
This work presents an event-triggered switching control framework for a class of nonlinear underactuated multi-channel systems with input constraints. These systems are inspired by cooperative manipulation tasks involving underactuation, where multiple underactuated agents collaboratively push or pull an object to a target pose. Unlike existing approaches for multi-channel systems, our method addresses underactuation and the potential loss of controllability by additionally addressing channel assignment of agents. To simultaneously account for channel assignment, input constraints, and stabilization, we formulate the control problem as a Mixed Integer Linear Programming and derive sufficient conditions for its feasibility. To improve real-time computation efficiency, we introduce an event-triggered control scheme that maintains stability even between switching events through a quadratic programming-based stabilizing controller. \blue{We theoretically establish the semi-global exponential stability of the proposed method and the asymptotic stability of its extension to nonprehensile cooperative manipulation under noninstantaneous switching. The proposed framework is further validated through numerical simulations on 2D and 3D free-flyer systems and multi-robot nonprehensile pushing tasks.
}

\end{abstract}


\section{Introduction}

\begin{figure}
    \centering
    \includegraphics[width=1.0\linewidth]{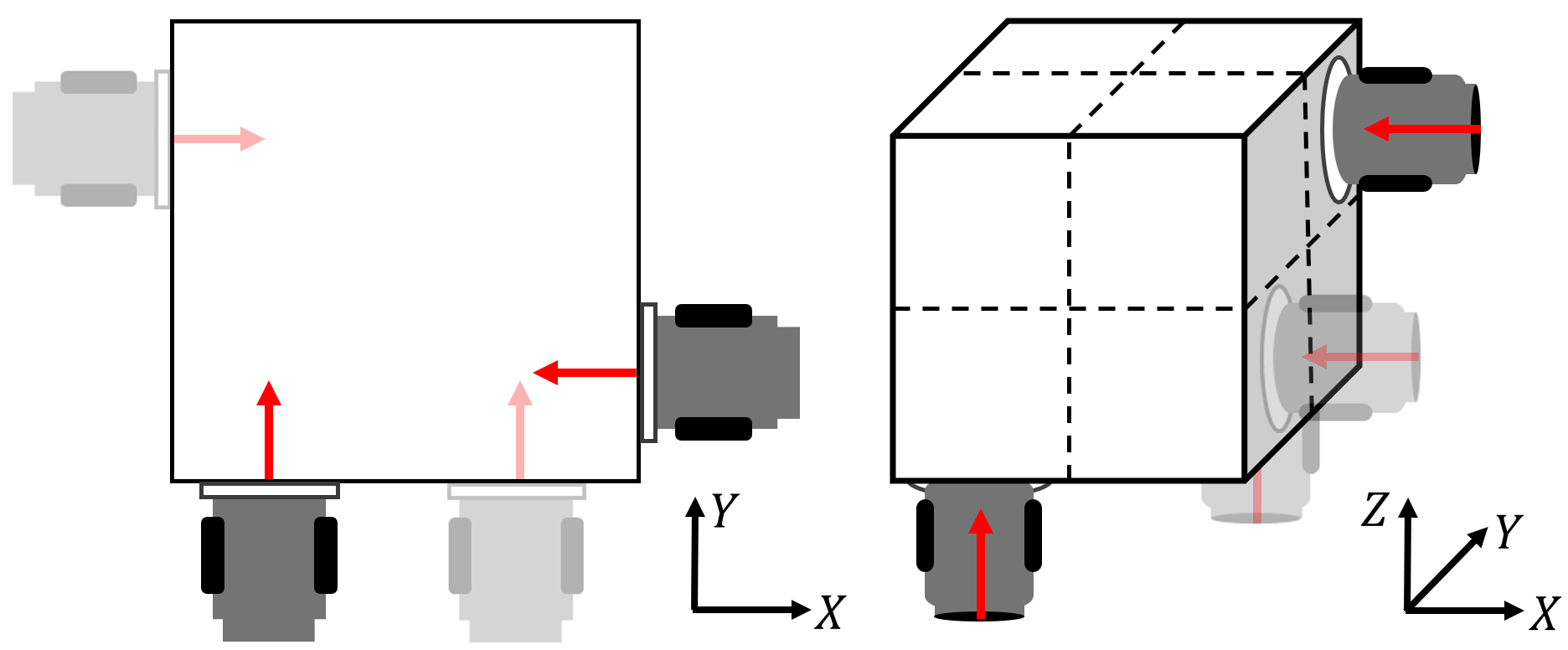}
    \caption{Two agents performing cooperative manipulation of a planar object (left) and a spatial object (right). The transparency of agents indicates the agents' pose before contact switching, and the red line is for the force exerted to the object.}
    \label{fig:scenario}
\end{figure}

Cooperative tasks involving objects that are collectively controlled by multiple agents such as drone swarms and robotic arms in manufacturing rely on precise object manipulation. However, when multiple agents with limited actuation control an object, ensuring coordinated motion becomes a significant challenge. This problem can be modeled as a multi-channel systems, and to broaden the scope of such cooperative control, this study investigates a class of nonlinear, underactuated multi-channel systems. In particular, we focus on a cooperative manipulation scenario where, unlike in \cite{culbertson2021decentralized, franchi2019distributed}, the object's configuration cannot be fully actuated by the control inputs provided by each agent (Fig. \ref{fig:scenario}). Such a scenario includes cases where each agent applies a one-dimensional thrust force \cite{lynch1999controllability} or where the control input is constrained to one or two dimensions, depending on the presence of friction, as in nonprehensile pushing manipulation \cite{hogan2020reactive}.

The challenge of underactuated agents in multi-agent cooperation has been discussed in the context of multi-channel systems \cite{kim2024decentralized, duan2023distributed}. These studies have explored distributed control for linear multi-channel systems, assuming that the overall system remains jointly controllable from the perspective of multiple agents. In particular, \cite{kim2024decentralized} did not consider the assignment of agents to specific control channels as a control variable, assuming that the system remains jointly controllable regardless of the channel through which each agent participates. \blue{However, the existing methodologies become inapplicable if 1) nonlinearities exist in the system (e.g., due to the rotational dynamics of the manipulated object in cooperative object manipulation), or 2) the number of participating agents is insufficient from the outset, or 3) some agents fail during the task, leading to a loss of joint controllability.}


This study seeks to develop a control approach for multi-channel systems without the assumptions of system linearity and joint controllability considered in previous studies. Specifically, we examine a scenario where the entire multi-agent system is jointly underactuated. Considering cooperative manipulation, this approach not only enables a comprehensive consideration of the rotational dynamics of the manipulated object but also accounts for underactuation in scenarios where the number of participating agents is limited. To address the challenges arising from this underactuation and the potential loss of controllability, unlike the approach in \cite{kim2024decentralized}, we treat not only the control inputs generated by participating agents but also the assignment of agents to specific channels as control variables. To this end, we propose a suitable switching controller for a class of nonlinear underactuated multi-channel systems.

In particular, we focus on real-time control, presenting a method that guarantees goal convergence while considering switching. The system that we consider has constraints on the input direction (analogous to allowing only pushing forces) and an upper bound on the magnitude of the input, reflecting the constraints found in real-world systems. The proposed framework, motivated by cooperative manipulation, can also be applied to nonprehensile pushing manipulation with minor modifications and can represent cooperative transportation with robotic platforms equipped with thruster-based actuators, such as Astrobee \cite{bualat2015astrobee} and microgravity free-flyers \cite{nakka2018six, pedro2024predictive}.

In the robotics community, similar challenges in underactuated cooperative manipulation have been addressed in the context of nonprehensile pushing manipulation, where a single agent, in most cases, manipulates a planar object using only pushing forces. This task inherently involves underactuation \cite{hogan2020reactive} and requires additional considerations due to the limitation that the agent can only push and not pull. A natural solution to address this underactuation and directional input constraint is to change the contact position and push the object from different locations. This requires solving a problem that involves both discrete and continuous variables to determine the switching of contact points and the pushing forces applied afterward.

To address this problem, \blue{the authors in \cite{hogan2020reactive}} proposed a model predictive control approach based on Mixed Integer Programming (MIP), while incorporating supervised learning to improve computational efficiency in determining the switching sequence. \blue{The authors in \cite{doshi2020hybrid}} introduced hybrid differential dynamic programming (DDP) that accounts for the hybrid dynamics involved. \blue{The work in \cite{xue2023guided}}, motivated by the observation that humans naturally excel at making discrete decisions, utilized human demonstrations as guidance for solving DDP. Furthermore, \blue{the authors of \cite{chi2024diffusion}} proposed a method that generates receding-horizon trajectories using a diffusion model trained on demonstration data. However, these approaches do not guarantee goal convergence of the manipulated object, resulting in a nonzero failure rate. Meanwhile, \blue{the study in \cite{graesdal2024towards}} approached the problem from a global planning perspective by formulating the nonconvex hybrid optimization problem as a convex problem through relaxation, achieving the empirical success rate of 100$\%$. However, the approach involves relatively high computational costs, which limits its applicability to real-time adaptation, and its theoretical performance guarantees remain unclear.

Recently, nonprehensile manipulation has been tackled in multi-agent settings. \blue{The authors in \cite{feng2024learning}} investigated collaborative pushing using two quadrupedal robots in an obstacle-filled environment, yet the approach did not provide a formal guarantee for goal convergence. \blue{The authors of \cite{tang2024collaborative}} proposed an algorithm with completeness guarantees under the assumptions of constant velocity transitions and quasi-static states. However, the method is applicable only in offline settings, highlighting the need for real-time decision-making methods that determine switching sequences online.

The contributions of this paper can be thus summarized as follows:
\begin{itemize}
    \item We propose a switching control framework for a class of nonlinear underactuated multi-channel systems. The controller is formulated as a Mixed Integer Linear Programming (MILP) problem, and we derive a sufficient condition to guarantee the feasibility of the proposed optimization under constraints on input direction and input magnitude (Theorem \ref{theorem - feasibility}).
    \item To improve the real-time computation of the proposed MILP-based controller, we introduce an event-triggered control scheme. Unlike conventional event-triggered control approaches, our method continuously considers input direction constraints between event triggers by designing a stabilizing quadratic programming (QP)-based controller, whose closed-form solution is derived (Theorem \ref{theorem: QP analytic sol}).
    \item We analyze the semi-global exponential stability of the proposed event-triggered switching control method (Theorem \ref{theorem: stability}) and validate our approach through numerical simulations on 2D and 3D free-flyer systems.
    \item \blue{We further demonstrate the applicability of the proposed methodology to nonprehensile cooperative manipulation, providing both theoretical analysis (Theorem~\ref{theorem - with delay}) and simulation validation under noninstantaneous switching.}
\end{itemize}

The structure of this paper is as follows. Section \ref{section-problem formulation} presents the problem formulation, including the system model and constraints considered. Section \ref{section-event-triggered switching control} introduces the proposed event-triggered switching controller, analyzes the feasibility of the constrained optimization-based controller, and provides a stability analysis of the closed-loop system. \blue{In Section~\ref{section: application to cooperative manipuulation}, the proposed controller is applied to the nonprehensile cooperative manipulation problem, and the case involving noninstantaneous switching is analyzed.} In Section \ref{section-simulation results}, we validate the proposed control framework through numerical simulations on 2D and 3D free-flyer systems. Finally, concluding remarks are provided in Section \ref{section-conclusion}.

\section{Problem Formulation} \label{section-problem formulation}

\subsection{Notations and preliminaries}
$[N]$ denotes the set $\{1,2,\cdots, N \}$ for an integer $N \geq 1$. For a vector $s \in \mathbb{R}^n$, $s_i$ is the $i^{th}$ element of $s$, and $\lVert s \rVert_\infty = \max_i \lvert s_i \rvert$. If not specified, the norm of a vector or matrix refers to the 2-norm or the induced 2-norm. The column space of a matrix $A$ is denoted as $\mathcal{C}(A)$. For two column vectors $a \in \mathbb{R}^m, b \in \mathbb{R}^n$, we define $[a;b] = [a^\top \ b^\top]^\top$. $1_n = [1;1;\cdots;1] \in \mathbb{R}^n$ and $0_n = [0;0;\cdots;0] \in \mathbb{R}^n$. $I_n$ is $n \times n$ identity matrix. The minimum and the maximum eigenvalue of a symmetric, positive-definite matrix $A$ is denoted as $\lambda_m(A)$ and $\lambda_M(A)$, respectively. We note the binary set as $\mathbb{Z}_2 = \{0,1 \}$, and use $\mathbb{Z}_2^n = \mathbb{Z}_2 \times \cdots \times \mathbb{Z}_2$ for its $n$-times Cartesian product. 

\begin{definition}[positive span\cite{regis2016properties}] \label{def-positive span}
    The positive span of a finite set of vectors $S = \{v_1, \cdots, v_k \} \subset \mathbb{R}^n$, denoted by $\text{pos}(S)$, is given by $\text{pos}(S) = \{\lambda_1 v_1 + \cdots \lambda_k v_k: \lambda_i \geq 0 \ \forall i = 1,\cdots,k \}$.
    The set $S$ is said to positively span $\mathbb{R}^n$ if $\text{pos}(S) = \mathbb{R}^n$.
\end{definition}

\subsection{System model}

For configuration $q \in \mathbb{R}^n$, we consider the following multi-channel system:
\begin{equation} \label{eq: eqn of motion}
\begin{aligned}
    \ddot{q} &= f(q,\dot{q}) + G(q,\dot{q})
    \sum_{i=1}^m b_i \delta_i u_i \\
    &= f(q,\dot{q}) + G(q,\dot{q}) B(\delta) u,
\end{aligned}
\end{equation}
where $u = \begin{bmatrix} u_1; \cdots; u_m \end{bmatrix} \in \mathbb{R}^{m}$, $\delta = \begin{bmatrix} \delta_1; \cdots; \delta_m \end{bmatrix} \in \mathbb{Z}_2^m$ and $B(\delta) = \begin{bmatrix} b_1 \delta_1, \cdots, b_m \delta_m \end{bmatrix} \in \mathbb{R}^{n \times m}$. Here, $b_i \in \mathbb{R}^n$ $\forall i \in [m]$ is a constant vector. The control inputs are $u$ and $\delta$, where $\delta_i \in \{0,1 \}$'s are indicator variables showing which input channel is activated. For a mechanical system, Coriolis-centrifugal force, friction, and gravitational force can be captured in $f(q,\dot{q})$. If not ambiguous, we use $B = B(1_m)$ for simplicity.

The objective is to design a tracking controller for $(u,\delta)$ where only up to $n_a \in [m]$ number of input channels can be activated at any instance while switching among the input channels is allowed. This constraint can be equivalently set as $\sum_{i=1}^m \delta_i = n_a$. As additional requirements, we impose an input bound $0 \leq u \leq u_u$ where $u_u >0$ is a constant. This problem setup can be summarized as follows:
\begin{problem} \label{problem}
    For given $n_a,u_u>0$, $q_d(t) \in \mathrm{C}^2$ and the system (\ref{eq: eqn of motion}), design a tracking controller for $(u,\delta)$ such that $\underset{t \to \infty}{\lim}q_d(t) - q(t) = 0$ while satisfying $\sum_{i=1}^m \delta_i = n_a$ and $0 \leq u \leq u_u$.
\end{problem}
In this work, we consider the following assumptions:
\begin{assumption} \label{assumption - positive span}
    $\mathcal{C}(B)$ positively spans $\mathbb{R}^n$.
\end{assumption}
\begin{assumption} \label{assumption - system matrix}
    $f(q,\dot{q})$ and $G(q,\dot{q})$ are Lipschitz continuous, and $G(q,\dot{q})$ is invertible.
\end{assumption}
\begin{assumption} \label{assumption-qd}
    The desired trajectory $q_d(t)$ is at least twice differentiable, and $\ddot{q}_d(t)$ is continuous. Furthermore, $\ddot{q}_d(t)$ is uniformly bounded by $\lVert \ddot{q}_d(t) \rVert_{\infty} \leq h_a$ with a positive constant $h_a$, and the bound is known.
\end{assumption}
\noindent Note that many of the applications of the considered multi-channel system including cooperative load transportation \cite{kim2024decentralized} and free-flyer robots actuated by propulsion systems \cite{daley2020astrobee}, \cite[Ch. 10]{pedro2024predictive} satisfy Assumptions \ref{assumption - positive span} and \ref{assumption - system matrix}. 
To distinguish activated input channels from their inactive counterparts, a selection matrix $E(\delta) \in \mathbb{R}^{n_a \times m}$ is defined as follows:
 \begin{equation} \label{eq: selection matrix}
     E_{i,j} = \begin{cases}
         1 & \text{if } i=j \wedge \delta_i = 1 \\
         0 & \text{otherwise}
     \end{cases}, 
 \end{equation}
 where $E_{i,j}$ is the $(i,j)$ elements of $E(\delta)$. 


\section{Event-Triggered Switching Control} \label{section-event-triggered switching control}

\begin{figure}
    \centering
    \includegraphics[width=1.0\linewidth]{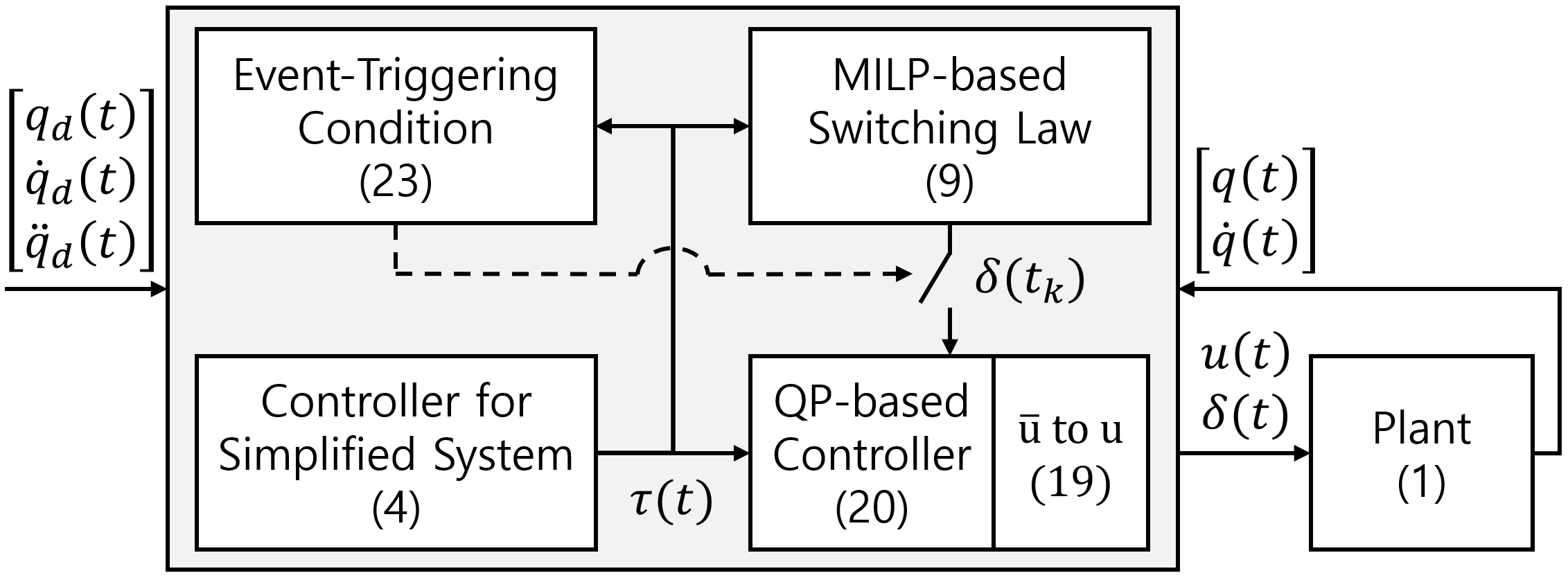}
    \caption{Flow chart of the overall control law.}
    \label{fig:flowchart}
\end{figure}

The overall flowchart of the proposed control law is summarized in Fig. \ref{fig:flowchart}. First, we construct a stabilizing controller for a simplified system and a Lyapunov function that corresponds to such system. Then, the time derivative of the Lyapunov function with respect to the actual system is analyzed. Next, we propose a switching controller using Mixed Inter Programming (MIP) that ensures a monotonic decrease of the Lyapunov function for the actual system. Lastly, an event-triggering condition and a QP-based controller are introduced in order to alleviate the computational burden of MIP. 

\subsection{Stabilizing controller for a simplified system}

Let us first consider the following simplified system:
\begin{equation} \label{eq: eqn of motion - simplified}
    \ddot{q} = f(q,\dot{q}) + G(q,\dot{q}) \tau,
\end{equation}
where $\tau \in \mathbb{R}^n$ substitutes $B(\delta) u$ in (\ref{eq: eqn of motion}). In the sequel, for notational simplicity, we denote $f(q,\dot{q}), G(q,\dot{q})$ without arguments unless they are ambiguous. A stabilizing controller can be designed for $\tau$ using a backstepping-like controller as follows:
\begin{equation} \label{eq: backstepping ctrller}
\begin{aligned}
    v_d &= \dot{q}_d + K_q (q_d - q), \\
    \tau &= G^{-1} \left( -f + \dot{v}_d + K_v (v_d - v) \right),
\end{aligned}
\end{equation}
\blue{where $v = \dot{q}$, and $q_d, v_d$ are the desired trajectories of $q$ and $v$, which are assumed to be sufficiently smooth as in Assumption \ref{assumption-qd}.} Defining $e_q = q_d - q$ and $e_v = v_d - v$, exponential stability of the closed-loop system composed of (\ref{eq: eqn of motion - simplified}) and (\ref{eq: backstepping ctrller}) can be easily verifed using a candidate Lyapunov function $V = \frac{1}{2}(e_q^\top e_q + e_v^\top e_v)$ if $K_q, K_v > \frac{1}{2} I_n$.

The closed-loop error dynamics of the actual system can be computed from (\ref{eq: eqn of motion}) and (\ref{eq: backstepping ctrller}) by adding and subtracting $G\tau$ to the RHS of (\ref{eq: eqn of motion}) as
\begin{equation} \label{eq: closed-loop system - actual}
\begin{aligned}
    \dot{e}_q &= -K_q e_q + e_v, \\
    \dot{e}_v &= -K_v e_v + G(\tau - B(\delta)u).
\end{aligned}
\end{equation}

Thus, the time-derivative of $V$ for the actual closed-loop system (\ref{eq: closed-loop system - actual}) can be computed as
\begin{equation} \label{eq: Vdot analysis}
\begin{aligned}
    \dot{V} &= -e_q^\top K_q e_q + e_q^\top e_v -e_v^\top K_v e_v + e_v^\top G (\tau - B(\delta)u) \\
    &\leq -cV + e_v^\top G (\tau - B(\delta)u),
\end{aligned}
\end{equation}
where $c = \min\{2 \lambda_m(K_q) - 1, 2 \lambda_m(K_v) - 1\} > 0$.

\subsection{Switching controller using MIP}
\label{subsec: switching law using MIP}

In realizing the computed generalized torque $\tau$ (\ref{eq: backstepping ctrller}), both $u$ and $\delta$ should be determined. 
We formulate an Mixed Integer Program for a switching controller as follows:
\begin{subequations} \label{eq: switching controller}
\begin{align}
    \min_{u,\delta,\rho} & & & -\rho \label{eq: ICS-cost} \\
    \text{s.t.} & & & 0 \leq u \leq u_u, \label{eq: ICS-pushing const} \\
    & & & 0 \leq \rho, \label{eq: ICS-nonZeno term} \\
    & & & -c_d V \geq -cV + e_v^\top G (\tau - B(\delta)u) + \rho, \label{eq: ICS-stability const} \\
    & & & \sum_{i\in [m]} \delta_i = n_a, \label{eq: ICS-agent number const}
\end{align}
\end{subequations}
where we introduce an auxiliary variable $\rho$ with a constraint to be non-negative (\ref{eq: ICS-nonZeno term}) to find a solution that minimizes $\dot{V}$. 
The input bound constraint (\ref{eq: ICS-pushing const}), constraint regarding stability (\ref{eq: ICS-stability const}), and number of active input channels (\ref{eq: ICS-agent number const}) are imposed. $c_d \in (0, c)$ is a constant parameter defining the decay rate of the Lyapunov function $V$.
As will be discussed in Theorem \ref{theorem - feasibility}, there always exist feasible $u, \delta, \rho$ satisfying all the constraints in (\ref{eq: switching controller}). Furthermore, we can always guarantee that $\rho > 0$ if $V > 0$.

Recalling that $B(\delta) u = \sum^m_{i=1} b_i \delta_i u_i$, due to the multiplicative term $u_i \delta_i$ in $B(\delta) u$, (\ref{eq: switching controller}) itself is not MILP. Thus, the current form cannot be solved using off-the-shelf optimization solvers like CPLEX \cite{cplex} or GUROBI \cite{gurobi}. To resolve this issue, we introduce an auxiliary variable $z_i$ $\forall i\in [m]$ with the following additional inequality constraints:
\begin{equation} \label{eq: big M method}
\begin{gathered}
    z_i \leq u_i, \quad u_u (1- \delta_i) + z_i \geq u_i \\
    -u_u \delta_i + z_i \leq 0, \quad z_i \geq 0.
\end{gathered}
\end{equation}
These inequality constraints allow us to replace $u_i \delta_i$ with $z_i$ \cite[Ch. 16.4]{borrelli2017predictive}, and the term $B(\delta) u$ can be rewritten as $B(\delta) u = \sum_i b_i z_i = B z$ where $z = [z_1;\cdots;z_m] \in \mathbb{R}^m$.

Using (\ref{eq: big M method}), (\ref{eq: switching controller}) can be translated into the following MILP problem:
\begin{equation} \label{eq: switching controller - MILP}
\begin{aligned}
    \min_{u,\delta,z,\rho} & & & -\rho \\
    \text{s.t.} & & & (\ref{eq: big M method}), \quad 0 \leq \rho, \\
    & & & -c_d V \geq -cV + e_v^\top G(\tau - B z) + \rho, \\    
    & & & \sum_i \delta_i = n_a,
\end{aligned}
\end{equation}
where the constraint on $u$, i.e. $0 \leq u \leq u_u$, is omitted as such constraint is included in (\ref{eq: big M method}).

For the ease of notation in the following feasibility analysis of (\ref{eq: switching controller}), we define $s_b,r$ as follows:
\begin{equation} \label{eq: sb, r}
    \quad s_b =G^\top e_v, \quad r= -(c-c_d)V + e_v^\top G \tau.
\end{equation}
The stability constraint (\ref{eq: ICS-stability const}) then can be rewritten as
\begin{equation} \label{eq: stability const - modified}
    r + \rho \leq s_b^\top B(\delta) u.
\end{equation}
\blue{Before proving in Theorem~\ref{theorem - feasibility} that a feasible solution to the proposed MIP (\ref{eq: switching controller}) always exists when the upper bound $u_u$ is chosen sufficiently large, we first establish Lemma~\ref{lemma - d, d1} by utilizing the property of Assumption~\ref{assumption - positive span}. Intuitively speaking, Lemma~\ref{lemma - d, d1} states that for a matrix $B$ satisfying Assumption~\ref{assumption - positive span}, there always exists at least one column vector of $B$ that has a positive inner product with an arbitrarily given vector $v$, and that the minimum magnitude of such an inner product is strictly positive.}
\begin{lemma} \label{lemma - d, d1}
    Let Assumption \ref{assumption - positive span} hold and $\max^{k}(A)$ indicate a set that has the $k$ largest elements of the set $A$. For $S(e) = \textstyle\max^{n_a} \{ m(b_1^\top e), \cdots, m(b_m^\top e) \}$ where $m(a) = \max\{0,a\}$, define 
    \begin{equation} \label{eq: d, d1}
    \begin{aligned}
        d_1 &=\min_{\lVert e \rVert = 1} \left( \max_{b_i \in \mathcal{C}(B)} b_i^\top e \right), \\
        d &= \min_{\lVert e \rVert = 1} \sum_{s \in S(e)} s > 0.
    \end{aligned}
    \end{equation}
    Then, $d, d_1$ satisfy the following:
    \begin{equation} \label{eq: d}
        d \geq d_1 > 0.
    \end{equation}
\end{lemma}
\begin{proof}
    We first show that $d_1 > 0$. Assume that $d_1 \leq 0$. Then, there exists $\lVert e \rVert = 1$ such that $\max_{b_i \in \mathcal{C}(B)} b_i^\top e \leq 0$. However, from \cite[Thm. 2.6]{regis2016properties} and Assumption \ref{assumption - positive span}, there always exist $b_i \in \mathcal{C}(B)$ such that $b_i^\top e > 0$ for every $e \in \mathbb{R}^n$. Thus, $d_1 > 0$ by contradiction.
    
    For any $e \in \mathbb{R}^n$ satisfying $\lVert e \rVert = 1$, the following holds by the definition of $m(\cdot)$ and the fact that $d_1 > 0$:
    \begin{equation*}
        \sum_{s \in S(e)} s \geq \textstyle\max \{ m(b_1^\top e), \cdots, m(b_m^\top e) \} = \max_{b_i \in \mathcal{C}(B)} b_i^\top e.
    \end{equation*}
    Thus, $d \geq d_1 > 0$, and this completes the proof.
\end{proof}


\begin{theorem} \label{theorem - feasibility}
    Let Assumptions \ref{assumption - positive span}, \ref{assumption - system matrix} hold and assume the following:
    \begin{equation} \label{eq: upper bound condition}
        u_u \geq \frac{ h + \left(\lambda_M(K_q + K_v) + \lambda_M(K_q^2) \right)\sqrt{2 V} }{d \sqrt{\lambda_m(GG^\top)} } = \mu(t),
    \end{equation}
    where $h = \lVert f(q,\dot{q}) \rVert + \lVert \ddot{q}_d \rVert$ and $d$ is defined in (\ref{eq: d, d1}). Then, for any $n_a \geq 1$, there exists a tuple of $(\delta,u,\rho)$ satisfying (\ref{eq: ICS-pushing const}), (\ref{eq: ICS-nonZeno term}), (\ref{eq: ICS-stability const}) and (\ref{eq: ICS-agent number const}). Furthermore, if $V > 0$, the solution satisfies $\rho > 0$.
\end{theorem}
\begin{proof}
    First, $\lambda_m(GG^\top) > 0$ since $G$ is invertible from Assumption \ref{assumption - system matrix}. If $s_b = 0$, or equivalently $e_v = 0$, then (\ref{eq: ICS-nonZeno term}) and (\ref{eq: ICS-stability const}) can be jointly written as $0 \leq \rho \leq (c-c_d) V$ by referring to (\ref{eq: stability const - modified}). In this case, because such constraint on $\rho$ and (\ref{eq: ICS-pushing const}) and (\ref{eq: ICS-agent number const}) are mutually disjoint, a feasible solution exists. Noting that the optimization problem (\ref{eq: switching controller}) maximizes $\rho$, the solution for $\rho$ can be obtained as $\rho = (c-c_d) V$, and $\rho > 0$ if $V > 0$ follows trivially. 

    Next, let us consider a case where $s_b \neq 0$. The following index set $J$ is considered:
    \begin{equation}
        J = \textstyle\argmax^{n_a} \{ m(b_1^\top s_b), \cdots, m(b_m^\top s_b) \},
    \end{equation}
    where $\argmax^k(A)$ denotes an index set that corresponds to the $k$ largest elements of the set $A$. Then, define $\delta, u$ as follows:
    \begin{equation} \label{eq: delta and u definition}
        \delta_i = \begin{cases}
            1 & i\in J \\
            0 & i\notin J
        \end{cases}, \quad 
        u_i = \begin{cases}
            u_u & i \in J \wedge b_i^\top s_b > 0 \\
            0 & i \notin J \vee  b_i^\top s_b \leq 0
        \end{cases}.
    \end{equation}
    From the definition of $B(\delta)$ and Lemma \ref{lemma - d, d1},
    \begin{equation}
        s_b^\top B(\delta) u = \lVert s_b \rVert \sum_{i \in J} \frac{m(s_b^\top b_i)}{\lVert s_b \rVert}  u_u \geq \lVert s_b \rVert u_u d,
    \end{equation}
    where $d$ is defined in (\ref{eq: d, d1}).
    Using (\ref{eq: upper bound condition}) and the facts that $\lVert e_q \rVert, \lVert e_v \rVert \leq \sqrt{2V}$ and $\lVert s_b \rVert = \sqrt{e_v^\top G G^\top e_v} \geq \sqrt{\lambda_{m}(G G^\top)} \lVert e_v \rVert$, we have
    \begin{equation*}
    \begin{aligned}
        u_u \lVert s_b \rVert d &\geq \lVert e_v \rVert \left( h + \left(\lambda_M(K_q + K_v) + \lambda_M(K_q^2) \right)\sqrt{2V} \right) \\
        &\geq e_v^\top \left( -f(q,\dot{q}) + \ddot{q}_d + (K_q + K_v) e_v - K_q^2 e_q \right) \\
        &= e_v^\top G \tau > r,
    \end{aligned}
    \end{equation*}
    where the last strict inequality comes from the fact that $V>0$ since $s_b \neq 0$. Thus, (\ref{eq: ICS-pushing const}), (\ref{eq: ICS-nonZeno term}), (\ref{eq: ICS-stability const}) and (\ref{eq: ICS-agent number const}) are always feasible. Next, since there exists a pair $(u,\delta)$ satisfying (\ref{eq: ICS-pushing const}), (\ref{eq: ICS-agent number const}) and $s_b^\top B(\delta) u > r$, there exists $\rho > 0$ such that (\ref{eq: stability const - modified}) holds. Since the problem (\ref{eq: switching controller}) maximizes $\rho$, the solution satisfies $\rho > 0$. This completes the proof. 
\end{proof}
    
\begin{remark}
    For the case when the input bound is symmetric, i.e. $-u_u \leq u \leq u_u$ instead of $0 \leq u \leq u_u$ in (\ref{eq: ICS-pushing const}), a switching controller can be similarly formulated as (\ref{eq: switching controller}) just by replacing (\ref{eq: ICS-pushing const}) to $-u_u \leq u \leq u_u$.
    Feasibility of this modified switching controller can be analyzed even with a weaker assumption than Assumption \ref{assumption - positive span}: \textit{$\mathcal{C}(B)$ linearly spans $\mathbb{R}^n$.} Such analysis can be found in Appendix A.
\end{remark}

\begin{claim} \label{claim: argmax}
    Let Assumptions \ref{assumption - positive span}, \ref{assumption - system matrix} hold and denote $\delta^*$ as the solution of (\ref{eq: switching controller}). Then, there exists an index $j \in J$ such that $\delta^*_j = 1$ where $J \subset [m]$ is an index set that satisfies the following:
    \begin{equation} \label{eq: bj}
        J = \left\{ j \in [m] \ | \ b_j = \argmax_{b_i \in \mathcal{C}(B)} b_i^\top s_b \right\}.
    \end{equation}
\end{claim}
\begin{proof}
    See Appendix B.
\end{proof}
\noindent \blue{Claim~\ref{claim: argmax} is later used in the proof of Theorem~\ref{theorem: stability}. This result can be intuitively understood by noting that, in the MIP~(\ref{eq: switching controller}), maximizing $\rho$ requires maximizing the right-hand side of constraint~(\ref{eq: stability const - modified}) (or equivalently, (\ref{eq: ICS-stability const})), which is the only constraint that limits the maximum value of $\rho$.}

\subsection{Event-triggering condition and real-time control strategy} \label{subsec - event-triggering condition}

Since the controller is formulated as MIP, the controller may suffer from computational burden. To mitigate this issue, we propose an event-triggered control strategy. Briefly speaking, the switching controller is updated only when an event is triggered, and in between the events, we fix $\delta$ and design a real time controller for $u$. By formulating such a real time controller as a QP problem, constraints with respect to $u$ can be directly considered in the controller. 

\blue{Since not all input channels are activated, the controller must be constructed by selecting only the control inputs $u$ corresponding to the activated channels. To this end, for given $\delta$, we use the selection matrix $E(\delta)$ defined in (\ref{eq: selection matrix}).} For such $E(\delta)$, let 
$\bar{B}_\delta = B(\delta) E(\delta)^\top \in \mathbb{R}^{n \times n_a}$ whose column vectors correspond only to the activated input channels. A QP-based controller is then introduced for an effective control input $\bar{u} \in \mathbb{R}^{n_a}$, where constraints on $u$ are imposed including monotonic decrease of the Lyapunov function and non-negativeness of $\bar{u}$. We recover $u$ from $\bar{u}$ by defining
\begin{equation} \label{eq: u from ubar}
    u = E(\delta)^\top \bar{u},    
\end{equation}
and $B(\delta) u = \bar{B}_\delta \bar{u}$ follows by definition. For $x = [q;\dot{q};q_d;\dot{q}_d;\ddot{q}_d]$, the proposed controller is as follows:

\begin{equation} \label{eq: qp formulation}
\begin{aligned}
    \bar{u}(x) = \argmin_{\bar{u}} & & & \frac{1}{2} \bar{u}^\top \bar{u} \\
    \text{s.t.} & & & -c_d V \geq -c V + e_v^\top G(\tau - \bar{B}_\delta \bar{u}),\\
    & & & \bar{u} \geq 0.
\end{aligned}
\end{equation}

Recalling $r$ in (\ref{eq: sb, r}) and defining $s$ as 
\begin{equation} \label{eq: s}
    s = \bar{B}_\delta^\top G^\top e_v \in \mathbb{R}^{n_a},    
\end{equation}
the QP-based controller (\ref{eq: qp formulation}) can be equivalently set as
\begin{subequations} \label{eq: qp formulation-rearranged}
\begin{align}
    \bar{u}(x) = \argmin_{\bar{u}} & & & \frac{1}{2} \bar{u}^\top \bar{u} \label{eq: qp - cost} \\
    \text{s.t.} & & & s(x)^\top \bar{u} - r(x) \geq 0, \label{eq: qp - stability const}\\
    & & & \bar{u} \geq 0. \label{eq: qp - positivity}
\end{align}
\end{subequations} 

Next, we define the event-triggering condition using the following conditions:
\begin{subequations} \label{eq: event-triggering condition}
\begin{gather}
     \left(r > 0 \right) \wedge  \left(\max_{i \in [n_a]} s_i \leq 0 \right), \label{eq: event-triggering condition - feas.} \\
     \max_{i \in [n_a]} \bar{u}_i > u_u. \label{eq: event-triggering condition - input bound}
\end{gather}
\end{subequations} 
The event is triggered when (\ref{eq: event-triggering condition - feas.}) or (\ref{eq: event-triggering condition - input bound}) is true.
These conditions are to prevent infeasibility of the QP problem (\ref{eq: qp formulation}) and also to abide by the input upper bound $u_u$. Thus, the switching controller (\ref{eq: switching controller}) is activated only if (\ref{eq: event-triggering condition - feas.}) or (\ref{eq: event-triggering condition - input bound}) is true.

\begin{theorem} \label{theorem: QP analytic sol}
    Assume that (\ref{eq: event-triggering condition - feas.}) does not hold. Then, the analytic solution of the QP problem (\ref{eq: qp formulation-rearranged}) can be obtained as
    \begin{equation} \label{eq: QP analytic sol}
        \bar{u}_i = 
        \begin{cases}
            \cfrac{m(s_i) r}{ \sum_{j\in [n_a]} m(s_j)^2} & \text{if } r > 0 \\
            0 & \text{otherwise}
        \end{cases}
        \quad \forall i \in [n_a],
    \end{equation}
    where $m(a) = \max\{0,a\}$ $\forall a \in \mathbb{R}$. Furthermore, (\ref{eq: event-triggering condition - feas.}) is necessary and sufficient for the infeasibility of the QP-based controller (\ref{eq: qp formulation-rearranged}).
\end{theorem}
\begin{proof}
    We first derive the analytic solution (\ref{eq: QP analytic sol}) and use this result to obtain the necessary and sufficient condition for the infeasibility.
    
    \textbf{Case 1:} When $r \leq 0$, (\ref{eq: QP analytic sol}) gives $\bar{u} = 0$ which is indeed the optimal solution for the unconstrained problem of (\ref{eq: qp formulation-rearranged}). Note that such solution candidate satisfies all the constraints; thus, it is the optimal solution to the problem.

    \textbf{Case 2:} Next, consider the case when $r > 0$. Since the problem (\ref{eq: qp formulation}) is QP, the following KKT conditions provide necessary and sufficient conditions for optimality \cite[Ch. 5.5.3]{boyd2004convex}: 
    \begin{subequations}
    \begin{align}
        0 &\leq \begin{bmatrix} s^\top \\ I_{n_a} \end{bmatrix} \bar{u} + \begin{bmatrix} -r \\ 0_{n_a} \end{bmatrix}, \label{eq: KKT - feas.} \\
        \bar{u} &= \begin{bmatrix} s & I_{n_a} \end{bmatrix} \lambda, \label{eq: KKT - stat.} \\
        0 &\leq \lambda, \label{eq: KKT - dual feas.} \\
        0 &= \lambda^\top \left( \begin{bmatrix} s^\top \\ I_{n_a} \end{bmatrix} \bar{u} + \begin{bmatrix} -r \\ 0_{n_a} \end{bmatrix} \right), \label{eq: KKT - comple.}
    \end{align}
    \end{subequations}
    where $\lambda = [\lambda_1;\cdots;\lambda_{n_a + 1}] \in \mathbb{R}^{n_a+1}$ is the Lagrange multiplier. From (\ref{eq: KKT - stat.}) and (\ref{eq: KKT - comple.}), we obtain
    \begin{subequations}
    \begin{align}
        \left\{ \textstyle \left(\sum_{i\in [n_a]} s_i^2 \right) \lambda_1 + \sum_{i\in [n_a]} s_i \lambda_{i+1} - r \right\} \lambda_1 &= 0, \label{eq: KKT combined - 1} \\
        \left(s_i \lambda_1 + \lambda_{i+1} \right) \lambda_{i+1} &= 0 \ \forall i \in [n_a]. \label{eq: KKT combined - 2}
    \end{align}
    \end{subequations} 
    

    \textbf{Case 2.1:} Let us first consider the case when $s_i \geq 0$ $\forall i$. Then, since $\lambda \geq 0$ by (\ref{eq: KKT - dual feas.}), $\lambda_{i} = 0$ $\forall i\geq 2$ from (\ref{eq: KKT combined - 2}). From (\ref{eq: KKT combined - 1}), if $\lambda_1 = 0$, then $\bar{u} = 0$ from (\ref{eq: KKT - stat.}), and infeasibility occurs from (\ref{eq: KKT - feas.}). Thus, $\lambda_1 \neq 0$, and from (\ref{eq: KKT combined - 1}) and (\ref{eq: KKT - stat.}),
    \begin{equation} \label{eq: step 1}
        \lambda_1 = \cfrac{r}{\sum_{i\in [n_a]} s_i^2}, \quad 
        \bar{u}_i = \cfrac{s_i r}{\sum_{i\in [n_a]} s_i^2}.
    \end{equation}
    Note that the denominator is always positive by the assumption of the theorem. The solution (\ref{eq: step 1}) is equivalent to (\ref{eq: QP analytic sol}) by referring to the fact that $m(a) = a$ if $a \geq 0$.

    \textbf{Case 2.2:} Next, consider the case when some of $s_i$'s are negative (but not all, by the assumption), and define an index set $I  \subsetneq [n_a]$ to denote indices for such $s_i$'s. 
    Following the same procedure as in the previous Case 2.1, it can be shown that $\lambda_1 \neq 0$. Having this in mind, we claim the following for this case:
    \begin{claim}
        The following holds for $\lambda_{i+1}$ $\forall i \in [n_a]$:
        \begin{equation*}
            \begin{cases}
                \lambda_{i+1} = 0 & \text{if } i \in [n_a]\backslash I \\
                s_i \lambda_1 + \lambda_{i+1} = 0 & \text{otherwise}
            \end{cases}.
        \end{equation*}
    \end{claim}
    \begin{proof}
        Since $s_i$'s $\forall i \in [n_a]\backslash I$ are non-negative, from (\ref{eq: KKT combined - 2}), $\lambda_{i+1} = 0$ $\forall i \in [n_a]\backslash I$. 
        Next, we show $s_i \lambda_1 + \lambda_{i+1}=0$ $\forall i \in I$ by contradiction. Assume that there exist a non-empty index set $\bar{I} \subset I$ such that $s_i \lambda_1 + \lambda_{i+1} \neq 0$ $\forall i \in \bar{I}$. Then, from (\ref{eq: KKT combined - 2}), $\lambda_{i+1} = 0$ $\forall i \in \bar{I}$, and (\ref{eq: KKT combined - 1}) can be rewritten as
        \begin{equation} \label{eq: KKT combined - 1 - rearranged}
            \left(\sum_{i\in [n_a]} s_i^2 \right) \lambda_1 + \sum_{i\in I \backslash \bar{I}} s_i \lambda_{i+1} = r, 
        \end{equation}
        where we also use the fact that $\lambda_{i+1} = 0$ for all $i \in [n_a] \backslash I$ and $\lambda_1 \neq 0$. 
        Next, if the set $I\backslash \bar{I}$ is empty, $\bar{u}_i$ can be obtained as $\bar{u}_i = s_i \lambda_1 = \frac{s_i r}{ \sum_{i \in [n_a]} s_i^2}$ $\forall i \in [n_a]$ from (\ref{eq: KKT - stat.}) and (\ref{eq: KKT combined - 1 - rearranged}); however, this violates (\ref{eq: KKT - feas.}) as $\bar{u} < 0$ $\forall i \in I$. Thus, $\tilde{n}$, the cardinality of the set $I \backslash \bar{I}$, is non-zero. 
        
        Combining (\ref{eq: KKT combined - 1 - rearranged}) with $s_i \lambda_1 + \lambda_{i+1} = 0$ $\forall i \in I \backslash \bar{I}$ yields
        \begin{equation} \label{eq: block matrix}
            \begin{bmatrix}
                \sum_{i\in [n_a]} s_i^2 & S^\top \\
                S & I_{\tilde{n}}
            \end{bmatrix}
            \begin{bmatrix}
                \lambda_1 \\
                \Lambda
            \end{bmatrix} = 
            \begin{bmatrix}
                r \\
                0_{\tilde{n}}
            \end{bmatrix},
        \end{equation}
        where $S, \Lambda \in \mathbb{R}^{\tilde{n}}$ are vectors that respectively concatenate $s_i$'s and $\lambda_{i+1}$'s $\forall i \in I \backslash \bar{I}$. The equation (\ref{eq: block matrix}) can be solved by applying block matrix inversion \cite{lu2002inverses} as
        \begin{equation} \label{eq: block matrix solved}
            \begin{bmatrix}
                \lambda_1 \\
                \Lambda
            \end{bmatrix} = \frac{1}{D} \begin{bmatrix}
                1 & -S^\top \\
                -S & D I_{\tilde{n}} + S S^\top
            \end{bmatrix} \begin{bmatrix}
                r \\ 0_{\tilde{n}}
            \end{bmatrix},
        \end{equation}
        where $D = \sum_{i \in [n_a]} s_i^2 - S^\top S = \sum_{i \in [n_a] \backslash (I \backslash \bar{I})} s_i^2$. Using (\ref{eq: block matrix solved}) and (\ref{eq: KKT - stat.}), we have
        \begin{equation*}
            \bar{u}_i = s_i \lambda_1 + \lambda_{i+1} = \frac{s_i r}{D} < 0 \ \forall i \in \bar{I},
        \end{equation*}
        which contradicts (\ref{eq: KKT - feas.}). Thus, $\bar{I}$ is an empty set, and this completes the proof.
    \end{proof}

    Based on Claim 2 and the proof therein, the KKT solution can be computed using (\ref{eq: KKT - stat.}) and (\ref{eq: block matrix solved}) with $\bar{I} = \emptyset$ as
    \begin{equation} \label{eq: ubar i when some s_i negative}
        \bar{u}_i = 
        \begin{cases}
            \cfrac{s_i r}{ \sum_{i \in [n_a] \backslash I} s_i^2} & \text{if } i \in [n_a] \backslash I \\
            0 & \text{otherwise}
        \end{cases}.
    \end{equation}
    Since $m(s_i) = 0$ for $i \in I$ and $m(s_i) = s_i$ for $i \in [n_a] \backslash I$, (\ref{eq: ubar i when some s_i negative}) can be compactly written as
    \begin{equation*}
        \bar{u}_i = \cfrac{m(s_i) r}{ \sum_{j\in [n_a]} m(s_j)^2}.
    \end{equation*}
    
    Next, we prove that (\ref{eq: event-triggering condition - feas.}) is necessary and sufficient for infeasibility of (\ref{eq: qp formulation-rearranged}). ($\Rightarrow$) The constraint on stability (\ref{eq: qp - stability const}) is infeasible in accordance with $\bar{u} \geq 0$ since $s^\top \bar{u} -r \leq -r < 0$. ($\Leftarrow$) In Theorem \ref{theorem: QP analytic sol}, we derive the closed-form feasible solution for the proposed QP, under the assumption that the condition (\ref{eq: event-triggering condition - feas.}) does not hold.
\end{proof}



\begin{remark} \label{remark - Lipschitz continuity}
    When (\ref{eq: qp formulation}) is feasible, the solution of (\ref{eq: qp formulation}), i.e. (\ref{eq: QP analytic sol}), is Lipschitz continuous for $x$. This can be shown by referring to \textit{Fact 1} and \textit{Fact 2} in \cite{xu2015robustness} and the fact that the function $m(\cdot)$ is Lipschitz.
\end{remark}

\begin{remark} \label{remark - continuity of solution}
    For $y=[q;\dot{q}]$, let us consider the following closed-loop system composed of dynamics (\ref{eq: eqn of motion}), QP-based controller (\ref{eq: QP analytic sol}), switching controller (\ref{eq: switching controller}) and the event-triggering condition (\ref{eq: event-triggering condition}):
    \begin{equation} \label{eq: switched system}
        \dot{y} = \begin{bmatrix}
            \dot{q} \\ f(q,\dot{q}) + G(q,\dot{q}) \bar{B}_\delta \bar{u}(x)
        \end{bmatrix} = f_{cl}(y,t).
    \end{equation}
    Such system (\ref{eq: switched system}) is a switched system \cite{liberzon2003switching} where switching occurs through $\delta$.
\end{remark}

\subsection{Stability analysis}

\begin{theorem} \label{theorem: stability}
    Let Assumptions \ref{assumption - positive span}, \ref{assumption - system matrix} and \ref{assumption-qd} hold and assume the existence of $h_b = \max\limits_{q_d(\cdot), \dot{q}_d(\cdot), V \leq V_0} \lVert f(q,\dot{q}) \rVert$ and the following: 
    \begin{equation} \label{eq: upper bound condition - constant}
        u_u \geq  \frac{ h_a + h_b + \left(\lambda_M(K_q + K_v) + \lambda_M(K_q^2) \right)\sqrt{2 V_0} }{d_1 \sqrt{\lambda_m(GG^\top)}} = \mu_{max},
    \end{equation}
    where $d_1$ is defined in (\ref{eq: d, d1}), $V_0 = V|_{t=0}$ and $h_a$ appears in Assumption \ref{assumption-qd}. Then, for any $n_a \geq 1$, the following statements hold:
    \begin{enumerate}[label=S\arabic*), ref=S\arabic*]
        \item $t_k < t_{k+1}$ $\forall k \in \mathbb{N}$ where $\{t_k\}_{k \in \mathbb{N}}$ is the sequence of event-triggered times. \label{thm3 - tk}
        \item The switching controller (\ref{eq: switching controller}) is feasible at $t_k$ $\forall k \in \mathbb{N}$. \label{thm3 - feasibility}
        \item The closed-loop error dynamics for $e_q$ and $e_v$ consisting of (\ref{eq: eqn of motion}), (\ref{eq: switching controller}), (\ref{eq: qp formulation}) and (\ref{eq: event-triggering condition}) is semiglobally exponentially stable. \label{thm3 - stability}
    \end{enumerate}
\end{theorem}
\begin{proof}
    The proof proceeds as follows. Assuming that $V(t_{k-1}) \leq V_0$ and $t_{k-1} < t_{k}$, we first show that the Lyapunov function $V$ exponentially decreases for the time interval $[t_{k-1}, t_{k})$ and thus $V(t_{k}) \leq e^{-c_d (t_{k}-t_{k-1})} V(t_{k-1})$ where $c_d > 0$ is the decay rate. Then, by using this property, we show that the premise of Theorem \ref{theorem - feasibility} holds, from which feasibility of the switching controller (\ref{eq: switching controller}) is guaranteed at $t=t_{k}$. Next, by showing that $\delta$ obtained from the switching controller (\ref{eq: switching controller}) and $\bar{u}$ from the QP-based controller (\ref{eq: QP analytic sol}) do not trigger the event (\ref{eq: event-triggering condition}), $t_{k} < t_{k+1}$ is derived. Lastly, we complete the proof by invoking mathematical induction.

    \textbf{Step 1:} During $t \in [t_{k-1}, t_{k})$, the QP-based controller (\ref{eq: qp formulation}) ensures that $\dot{V} \leq -c_d V$ from which we can obtain $V(t_{k}^-) \leq e^{-c_d (t_{k}-t_{k-1})} V(t_{k-1})$. Since the state $[q;\dot{q}]$ does not jump during switching considering the switched system (\ref{eq: switched system}) and since $q_d(t), \dot{q}_d(t), \ddot{q}(t)$ are continuous, $V(t)$ is continuous at $t = t_{k}$, i.e. $V(t_{k}) = V(t_{k}^-)$ and
    \begin{equation} \label{eq: V tk and tk-1}
        V(t_{k}) \leq e^{-c_d (t_{k}-t_{k-1})} V(t_{k-1}).
    \end{equation}

    \textbf{Step 2:} Now, recalling $\mu(t)$ defined in (\ref{eq: upper bound condition}),
    \begin{equation} \label{eq: mu bound}
    \begin{gathered}
     \mu(t_{k}) = \frac{ h(t_{k}) + \left(\lambda_M(K_q + K_v) + \lambda_M(K_q^2) \right)\sqrt{2 V(t_{k})} }{d \sqrt{\lambda_m(GG^\top)} }  \\
     \leq \frac{ h_a + h_b + \left(\lambda_M(K_q + K_v) + \lambda_M(K_q^2) \right)\sqrt{2 V_0} }{d_1 \sqrt{\lambda_m(GG^\top)} } = \mu_{max},
    \end{gathered}
    \end{equation}
    where we use $V(t_{k}) \leq V_0$, $0< d_1 < d$ from Lemma \ref{lemma - d, d1}, and $h(t_{k}) \leq h_a + \max_{q_d(\cdot),\dot{q}_d(\cdot),V\leq V(t_{k})} \lVert f(q,\dot{q}) \rVert \leq h_a + h_b$. From (\ref{eq: mu bound}), the premise of Theorem \ref{theorem - feasibility} holds, and thus the switching controller (\ref{eq: switching controller}) is feasible at $t= t_{k}$.

    \textbf{Step 3:} Next, we show $t_k < t_{k+1}$ by analyzing that the solution $\delta$ from the switching controller (\ref{eq: switching controller}) and $\bar{u}$ from the QP-based controller (\ref{eq: QP analytic sol}) do not trigger any of the two events in (\ref{eq: event-triggering condition}). For the optimal solution $u^*, \delta^*, \rho^*$ obtained from the switching controller (\ref{eq: switching controller}), by defining $\bar{u}^* = E(\delta^*) u^*$, $B(\delta^*) u^* = \bar{B}_{\delta^*} \bar{u}^*$ holds by definition. Then, the solution $u^*, \delta^*, \rho^*$ satisfies the following:
    \begin{equation} \label{eq: QP feasibility at switching}
    \begin{aligned}
        -c_d V &\geq -cV + e_v^\top G (\tau - B(\delta^*) u^*) +\rho^* \\
        &\geq -cV + e_v^\top G (\tau - B(\delta^*) u^*) \\
        &= -cV + e_v^\top G (\tau - \bar{B}_{\delta^*} \bar{u}^*),
    \end{aligned}
    \end{equation}
    where the inequalities directly come from the constraints (\ref{eq: ICS-stability const}) and (\ref{eq: ICS-nonZeno term}) imposed in the switching controller (\ref{eq: switching controller}). Furthermore, $\bar{u}^* \geq 0$ by (\ref{eq: ICS-pushing const}). Thus, $\bar{u}^*$ is a feasible solution to the QP-based controller (\ref{eq: qp formulation}), showing feasibility of (\ref{eq: qp formulation}). From Theorem \ref{theorem: QP analytic sol}, since (\ref{eq: qp formulation}) is feasible, (\ref{eq: event-triggering condition - feas.}) is not triggered.

    Now, let us show that (\ref{eq: event-triggering condition - input bound}) is also not triggered. If $\bar{u} = 0$ as in (\ref{eq: QP analytic sol}), then $\bar{u}_i = 0 \leq \mu_{max}$ $\forall i \in [n_a]$ is trivially true. Next, consider the other case in (\ref{eq: QP analytic sol}). Define $\mu_1$ as 
    \begin{equation} \label{eq: mu1}
        \mu_1 = \frac{d}{d_1} \mu,
    \end{equation}   
    where $\mu$ is defined in (\ref{eq: upper bound condition}), and $d, d_1$ appear in Lemma \ref{lemma - d, d1}. Referring to (\ref{eq: backstepping ctrller}) and (\ref{eq: upper bound condition}), 
    \begin{equation*}
        \lVert G \tau \rVert \leq \sqrt{\lambda_m(G G^\top)} d_1 \mu_1.
    \end{equation*}
    Accordingly, $r$ in (\ref{eq: sb, r}) satisfies the following inequalities:
    \begin{equation} \label{eq: m(r)}
    \begin{gathered}
        r \leq e_v^\top G \tau \leq | e_v^\top G \tau | = | s_b^\top G^{-1} G\tau | \leq \\
        \lVert s_b \rVert \lVert G^{-1} \rVert \lVert G \tau \rVert \leq \lVert s_b \rVert d_1 \mu_1,
    \end{gathered}
    \end{equation}
    where we use the fact that $\lVert G^{-1} \rVert = \frac{1}{\sqrt{\lambda_m(G G^\top)}}$ and $V\geq 0$. 
    Then,
    \begin{equation} \label{eq: sb d1}
        \lVert s_b \rVert d_1 \leq \max_{b_i \in \mathcal{C(B)}} b_i^\top s_b = \max_{j\in[n_a]} m(s_j),
    \end{equation}
    where the inequality is derived by the definition of $d_1$ in Lemma \ref{lemma - d, d1}, and the equality comes from the definition of $s$ in (\ref{eq: s}), Claim \ref{claim: argmax} and \cite[Thm. 2.6]{regis2016properties} implying $\max_{j\in[n_a]} s_j > 0$. Finally, combining (\ref{eq: m(r)}) and (\ref{eq: sb d1}), the following holds for all $i \in [n_a]$:
    \begin{equation} \label{eq: ubar mu1}
        \bar{u}_i = \frac{m(s_i) r}{\sum_{j \in [n_a]} m(s_j)^2} \leq \mu_1 \frac{m(s_i) \{ \max_{j\in[n_a]} m(s_j) \} }{\sum_{j \in [n_a]} m(s_j)^2} \leq \mu_1.
    \end{equation}
    Using (\ref{eq: upper bound condition - constant}), (\ref{eq: mu bound}), (\ref{eq: mu1}) and (\ref{eq: ubar mu1}) at $t_k$, $\bar{u}_i(t_k) \leq u_u$ $\forall i \in [n_a]$, which confirms that (\ref{eq: event-triggering condition - input bound}) is not triggered.

    Let $t_0 = 0$ and consider the two cases where $t_1 > t_0$ and $t_1 = t_0$. For the first case, $t_0 < t_1$ and $V(t_0) \leq V_0$ holds, and thus, by mathematical induction, the first two items \ref{thm3 - tk} and \ref{thm3 - feasibility} hold $\forall k \in \mathbb{N}$. Next, when $t_1 = t_0$, we can start from STEP 2 at $t_1$ since $V(t_1) = V(t_0) \leq V_0$. Then, from STEP 3, we can obtain $t_1 < t_2$, and by combining these results of $V(t_1) \leq V_0$ and $t_1 < t_2$, the first two statements \ref{thm3 - tk} and \ref{thm3 - feasibility} hold $\forall k \in \mathbb{N}\backslash \{1\}$ by mathematical induction.
    
    Lastly, since (\ref{eq: V tk and tk-1}) holds $\forall k \in \mathbb{N}$ and $\dot{V} \leq -c_d V$ $\forall t \in [t_k, t_{k+1})$ $\forall k \in \mathbb{N}$ by the QP-based controller (\ref{eq: qp formulation}),
    \begin{equation} \label{eq: V(t) exponential decrease}
        V(t) \leq e^{-c_d (t-t_k)} V(t_k) \leq e^{-c_d t} V_0 \quad \forall t \geq 0.
    \end{equation}
    Furthermore, since the input upper bound condition (\ref{eq: upper bound condition - constant}) applies only to the initial condition, exponential stability can be obtained for any given initial condition if the input bound is sufficiently large. This completes the proof for \ref{thm3 - stability}.
\end{proof}


\begin{remark}
    Considering a mechanical system, the term $f(q,\dot{q})$ consists of Coriolis-centrifugal force, friction and gravitational force. Note that these terms are mostly upper bounded in the sense of a vector norm if a norm of the generalized velocity, i.e. $\lVert \dot{q} \rVert$, is upper bounded. Thus, recalling $h_b = \max\limits_{q_d(\cdot), \dot{q}_d(\cdot), V \leq V_0} \lVert f(q,\dot{q}) \rVert$, finite $h_b$ exists if $\lVert \dot{q}_d(\cdot) \rVert$ is upper bounded.
\end{remark}

{\color{blue}
\section{Application to Nonprehensile Cooperative Manipulation} \label{section: application to cooperative manipuulation}

This section describes how the proposed methodology can be applied to the problem of nonprehensile cooperative manipulation. Specifically, we consider the scenario illustrated on the left of Fig.~\ref{fig:scenario}, where multiple robots collaboratively push a common planar object to move it toward a desired position and orientation. Such collaboration enables the robots to manipulate objects that are otherwise too heavy or large for a single robot, while also reducing individual effort. Following previous works~\cite{tang2024collaborative, hogan2020reactive, graesdal2024towards}, we consider the following quasi-static dynamics:
\begin{equation} \label{eq: eqn of motion for quasi-static}
    \dot{q} = 
    \begin{bmatrix}
        d_f^{-1} I_2 & 0_2 \\ 0_2^\top & d_\tau^{-1}
    \end{bmatrix}
    \begin{bmatrix}
        R & 0_2 \\ 0_2^\top & 1
    \end{bmatrix}
    B(\delta) u 
    =: G(q) B(\delta) u,
\end{equation}
where $d_f$ and $d_\tau$ are terms related to the static friction between the object and the surface, and whose exact definition is given in the following. The variables $u, \delta \in \mathbb{R}^m$ represent the pushing force exerted by each robot and the indicator variable specifying which contact channel each robot is attached to, respectively. More specifically, given the static friction coefficient $\mu$ between the object and the surface, the object mass $m$, gravitational acceleration $g$, characteristic distance $r$, and the geometry-dependent constant $c \in [0,1]$, we define $d_f = \mu m g$, $d_\tau = c r \mu m g$. More details of the derivation can be found in~\cite{graesdal2024towards, dafle2018inhand}. The matrix $B(\delta)$ is defined similarly to (\ref{eq: eqn of motion}), where its $i$-th column is $b_i \delta_i$. Here, $b_i \in \mathbb{R}^3$ denotes the mapping from the normal force applied at the $i$-th contact point to the corresponding force and torque on the object in the body frame.

Compared with (\ref{eq: eqn of motion}), the quasi-static dynamics in (\ref{eq: eqn of motion for quasi-static}) have an identical structure except that they are first-order instead of second-order. We first discuss how the original controller can be modified to account for this difference. Then, based on the modified controller, we analyze the stability of the system and the feasibility of the proposed optimization problem using results from the previous sections. Finally, we extend the analysis to address the practical case where switching cannot occur instantaneously in real robotic systems. Throughout this section, Assumptions~\ref{assumption - positive span} and~\ref{assumption - system matrix} are maintained, while Assumption~\ref{assumption-qd} is replaced by the following:
\begin{assumption} \label{assumption - qd qs}
    The desired trajectory $q_d(t)$ is differentiable, and $\dot{q}_d(t)$ is continuous. Furthermore, $\dot{q}_d(t)$ is uniformly bounded by $\lVert \dot{q}_d(t) \rVert_{\infty} \leq h_a$ with a known positive constant $h_a$.
\end{assumption}
\noindent Assumption \ref{assumption - qd qs} is essentially the same as Assumption \ref{assumption-qd}, reformulated to fit the new quasi-static system (\ref{eq: eqn of motion for quasi-static}).

\subsection{Controller Design Revisited}

The controller structure follows the same principle as the previously proposed method. Considering that the system is first-order, a simplified controller can be designed as
\begin{equation} \label{eq: simplified controller for qs}
    \tau = G^{-1} (\dot{q}_d + K_q e_q).
\end{equation}
We define a candidate Lyapunov function $V = \tfrac{1}{2} e_q^\top e_q$ in quadratic form. Then, the following inequality is derived:
\begin{equation} \label{eq: stability condition for quasi-static}
    \dot{V} \leq -cV + e_q^\top G(\tau - B(\delta)u),
\end{equation}
where $c = 2\lambda_m(K_q)$.

Similar to (\ref{eq: switching controller}), we can design a switching controller that considers (\ref{eq: stability condition for quasi-static}) instead of (\ref{eq: ICS-stability const}). The resulting formulation is as follows:
\begin{equation} \label{eq: switching controller - quasi-static}
\begin{aligned}
    \min_{u,\delta,\rho} & & & -\rho \\
    \text{s.t.} & & & 0 \leq u \leq u_u, \quad 0 \leq \rho, \\
    & & & -c_d V \geq -cV + e_q^\top G (\tau - B(\delta)u) + \rho, \\
    & & & \sum_{i\in [m]} \delta_i = n_a.
\end{aligned}
\end{equation}
Based on Theorem~\ref{theorem - feasibility}, we obtain the following corollary regarding the feasibility of (\ref{eq: switching controller - quasi-static}).

\begin{corollary}
\label{corollary - feasibility}
    Let Assumptions~\ref{assumption - positive span} and~\ref{assumption - system matrix} hold, and assume the following holds:
    \begin{equation} \label{eq: upper bound condition - qs}
        u_u \geq \frac{ \lVert \dot{q}_d \rVert + \lambda_M(K_q) \sqrt{2 V} }{d \sqrt{\lambda_m(GG^\top)} } = \mu(t),
    \end{equation}
    where $d$ is defined in (\ref{eq: d, d1}). Then, for any $n_a \geq 1$, there exists a tuple $(\delta,u,\rho)$ satisfying all constraints in (\ref{eq: switching controller - quasi-static}). Furthermore, if $V > 0$, the solution satisfies $\rho > 0$.
\end{corollary}
\begin{proof}
    The proof follows directly from the procedure in Theorem~\ref{theorem - feasibility}.
\end{proof}

The resulting MIP optimization problem can be converted into an MILP by introducing auxiliary variables, as in (\ref{eq: switching controller - MILP}). The event-triggering condition is defined identically to (\ref{eq: event-triggering condition}), and the QP-based controller follows the same structure as (\ref{eq: qp formulation-rearranged}). The variables $r$ and $s$ are defined analogously to (\ref{eq: sb, r}) and (\ref{eq: s}) as:
\begin{equation}\label{eq: s,r new}
    r = -(c-c_d)V + e_q^\top G \tau, \quad s = \bar{B}_\delta^\top G^\top e_q.
\end{equation}
The analytic solution of the QP can then be derived according to Theorem~\ref{theorem: QP analytic sol}.

As a corollary of Theorem~\ref{theorem: stability}, we analyze the stability of the closed-loop system as follows:
\begin{corollary}
\label{corollary: stability}
    Let Assumptions~\ref{assumption - positive span}, \ref{assumption - system matrix}, and \ref{assumption - qd qs} hold, and assume that
    \begin{equation} \label{eq: upper bound condition - constant - qs}
        u_u \geq  \frac{ h_a + \lambda_M(K_q) \sqrt{2 V_0} }{d_1 \sqrt{\lambda_m(GG^\top)}} = \mu_{\max},
    \end{equation}
    where $d_1$ is defined in (\ref{eq: d, d1}), $V_0 = V|_{t=0}$, and $h_a$ is given in Assumption~\ref{assumption - qd qs}. Then, for any $n_a \geq 1$, the following statements hold:
    \begin{enumerate}[label=S\arabic*), ref=S\arabic*]
        \item $t_k < t_{k+1}$ for all $k \in \mathbb{N}$, where $\{t_k\}_{k \in \mathbb{N}}$ is the sequence of event-triggered times. \label{cor2 - tk}
        \item The switching controller (\ref{eq: switching controller - quasi-static}) is feasible at $t_k$ $\forall k \in \mathbb{N}$. \label{cor2 - feasibility}
        \item The closed-loop error dynamics for $e_q$ consisting of (\ref{eq: eqn of motion for quasi-static}), (\ref{eq: switching controller - quasi-static}) and (\ref{eq: qp formulation-rearranged}), (\ref{eq: event-triggering condition}) with $r,s$ defined in (\ref{eq: s,r new}), are semiglobally exponentially stable. \label{cor2 - stability}
    \end{enumerate}
\end{corollary}
\begin{proof}
    The proof follows the same steps as Theorem~\ref{theorem: stability}.
\end{proof}

\subsection{Analysis for Noninstantaneous Switching}

In the above analysis, it was assumed that when an event trigger occurs and $\delta$ switches, the robot can immediately apply force through the newly activated channel. In practice, however, when $\delta$ is switched, each robot requires a finite amount of time to move to the new channel, during which no control input can be applied to the manipulated object. To incorporate such noninstantaneous switching behavior, we provide the following analysis under the additional condition that $\dot{q}_d = 0$. Such a condition, which is widely adopted in the nonprehensile manipulation literature \cite{xue2023guided, graesdal2024towards, chi2024diffusion}, can represent quasi-static settings where the desired configuration remains constant while the object is being manipulated toward the target.
\begin{theorem} \label{theorem - with delay}
    Let Assumptions~\ref{assumption - positive span} and~\ref{assumption - system matrix} hold, and assume that (\ref{eq: upper bound condition - constant - qs}) and $\dot{q}_d = 0$ hold. Let $\{t_k\}$ denote the sequence of event-triggered times, and define the control input $u$ based on the following:
    \begin{equation} \label{eq: QP control with delay}
        \bar{u}_i(t) = 
        \begin{cases}
            0, & \text{if } t \in [t_k, t_k + \Delta_k), \\
            (\ref{eq: QP analytic sol}), & \text{if } t \in [0, t_1) \cup [t_k + \Delta_k, t_{k+1}),
        \end{cases}
        \quad \forall i \in [n_a],
    \end{equation}
    where $k \in \mathbb{N}$, $r$ and $s$ in (\ref{eq: QP analytic sol}) are defined in (\ref{eq: s,r new}), $\Delta_k > 0$ denotes a finite time-delay term, and the relationship between $u$ and $\bar{u}$ follows (\ref{eq: u from ubar}). Then, for any $n_a \geq 1$, the following statements hold:
    \begin{enumerate}[label=S\arabic*), ref=S\arabic*]
        \item $t_{k+1} - t_k > \Delta_k$ for all $k \in \mathbb{N}$. \label{thm4 - tk}
        \item The switching controller (\ref{eq: switching controller - quasi-static}) is feasible at $t_k$ $\forall k \in \mathbb{N}$. \label{thm4 - feasibility}
        \item The closed-loop error dynamics for $e_q$, consisting of (\ref{eq: eqn of motion for quasi-static}), (\ref{eq: switching controller - quasi-static}), (\ref{eq: QP control with delay}), and (\ref{eq: event-triggering condition}) with $r,s$ defined in (\ref{eq: s,r new}), are semiglobally asymptotically stable. \label{thm4 - stability}
    \end{enumerate}
\end{theorem}
\begin{proof}
    Consider first the case with $\Delta_k = 0$ for all $k \in \mathbb{N}$. The corresponding closed-loop system is obtained by combining (\ref{eq: eqn of motion for quasi-static}), (\ref{eq: switching controller - quasi-static}), (\ref{eq: QP analytic sol}), and (\ref{eq: event-triggering condition}) with $r,s$ defined in (\ref{eq: s,r new}). Under the additional condition $\dot{q}_d(t) = 0$, this system can be written as an autonomous system~\cite[Ch.~4]{khalil2002nonlinear}:
    \begin{equation} \label{eq: closed-loop system - 1st order}
        \dot{e}_q(t) = f_{cl}(e_q(t); \delta(0), e_q(0)).
    \end{equation}

    Using (\ref{eq: closed-loop system - 1st order}) and by definition of (\ref{eq: QP control with delay}), the closed-loop system when $\Delta_k \neq 0$ described by
    (\ref{eq: eqn of motion for quasi-static}), (\ref{eq: switching controller - quasi-static}), (\ref{eq: QP control with delay}), and (\ref{eq: event-triggering condition}) with $r,s$ defined in (\ref{eq: s,r new}) becomes
    \begin{equation} \label{eq: closed-loop system - 1st order with delta}
        \dot{\tilde{e}}_q(t) =
        \begin{cases}
            0, & \text{if } t \in [t_k, t_k+\Delta_k), \\
            f_{cl}(\tilde{e}_q(t); \tilde{\delta}(t_k), \tilde{e}_q(t_k)), & \text{if } t \in [t_k+\Delta_k, t_{k+1}), \\
            f_{cl}(\tilde{e}_q(t); \tilde{\delta}(0), \tilde{e}_q(0)), & \text{if } t \in [0,t_1),
        \end{cases}
    \end{equation}
    where $\tilde{(\cdot)}$ denotes the variables of the system (\ref{eq: closed-loop system - 1st order with delta}) for distinction from (\ref{eq: closed-loop system - 1st order}). Similar to $\{t_k\}$, define $\{t_k'\}$ as the event-trigger time sequence for (\ref{eq: closed-loop system - 1st order}).

    \begin{claim} \label{claim 3}
        Suppose both closed-loop systems (\ref{eq: closed-loop system - 1st order}) and (\ref{eq: closed-loop system - 1st order with delta}) share the same initial conditions $\tilde{\delta}(0) = \delta(0)$ and $\tilde{e}_q(0) = e_q(0)$. Then, the following hold for $t_k$ and $\tilde{e}_q$:
        \begin{equation} \label{eq: t and t' relation - general}
            t_k = t_k' + \sum_{i\in [k-1]} \Delta_i,   
        \end{equation}
        \begin{equation} \label{eq: eq tilde and eq relation - general}
            \tilde{e}_q \! \left( \! t + \! \sum_{i \in [k-1]} \! \Delta_i \! \right) \! = \!
            \begin{cases}
               e_q(t_k'), & \text{if } t \in [t_k', t_k' + \Delta_k], \\ 
               e_q(t), &  \text{if } t \in [t_k' + \Delta_k, t_{k+1}' + \Delta_k).
            \end{cases}
        \end{equation}
    \end{claim}

    \begin{proof}
        Assume $\tilde{e}_q(t_k) = e_q(t_k')$ and $\tilde{\delta}(t_k) = \delta(t_k')$. Then (\ref{eq: closed-loop system - 1st order with delta}) implies $\tilde{e}_q(t_k + \Delta_k) = e_q(t_k')$ and $\tilde{\delta}(t_k + \Delta_k) = \delta(t_k')$. For $t \in [t_k + \Delta_k, t_{k+1})$, the vector fields of (\ref{eq: closed-loop system - 1st order}) and (\ref{eq: closed-loop system - 1st order with delta}) coincide and are autonomous. Hence, from the definition of an autonomous system \cite[Ch. 4]{khalil2002nonlinear}, the solution is time-translation invariant; thus,
        \begin{equation} \label{eq: eq tilde and eq relation}
            \tilde{e}_q(t) = e_q\big(t - t_k - \Delta_k + t_k'\big) \quad \forall t \in [t_k+\Delta_k, t_{k+1}).
        \end{equation}
        Moreover, since $r,s$ (\ref{eq: s,r new}) appearing in the event-trigger condition (\ref{eq: event-triggering condition}) are only functions of $e_q$ thanks to the condition $\dot{q}_d = 0$, by (\ref{eq: eq tilde and eq relation}), the event-trigger intervals must match:
        \begin{equation} \label{eq: t and t' relation}
            t_{k+1} - (t_k + \Delta_k) = t_{k+1}' - t_k'.
        \end{equation}
        Since both $e_q(t)$ and $\tilde{e}_q(t)$ are continuous, from (\ref{eq: eq tilde and eq relation}) and (\ref{eq: t and t' relation}), $\tilde{e}_q(t_{k+1}) = e_q(t'_{k+1})$. Furthermore, since the same switching controller (\ref{eq: switching controller - quasi-static}) is incorporated in both systems (\ref{eq: closed-loop system - 1st order}) and (\ref{eq: closed-loop system - 1st order with delta}) with the same state feedback, which are $e_{q}(t'_{k+1})$ and $\tilde{e}_q(t_{k+1})$, respectively, $\tilde{\delta}(t_{k+1}) = \delta(t'_{k+1})$.

        Depending on the initial value of $\tilde{\delta}(0)$, it is possible that either $t_1 = 0$ or $t_1 > 0$. When $t_1 = 0$, since the two systems share the same event-triggering condition, we have $t_1 = t_1'$, and because they employ the same switching controller (\ref{eq: switching controller - quasi-static}), it follows that $\tilde{\delta}(t_1) = \delta(t_1')$. On the other hand, if $t_1 > 0$, the closed-loop dynamics over $t \in [0, t_1)$ are identical in (\ref{eq: closed-loop system - 1st order}) and (\ref{eq: closed-loop system - 1st order with delta}), leading to the same event-triggering instant $t_1 = t_1'$ and $\tilde{e}_q(t_1) = e_q(t_1')$. Since the same switching controller is applied, it also follows that $\tilde{\delta}(t_1) = \delta(t_1')$. Therefore, regardless of whether $t_1 = 0$ or $t_1 > 0$, we have $t_1 = t_1'$, $\tilde{e}_q(t_1) = e_q(t_1')$, and $\tilde{\delta}(t_1) = \delta(t_1')$. 
        
        Thus, by mathematical induction, using the condition $t_1 = t_1'$ together with (\ref{eq: t and t' relation}), (\ref{eq: t and t' relation - general}) is obtained, and by combining (\ref{eq: t and t' relation - general}) with (\ref{eq: eq tilde and eq relation}), (\ref{eq: eq tilde and eq relation - general}) follows.
    \end{proof}

    Item~\ref{thm4 - tk} follows from (\ref{eq: t and t' relation - general}) in Claim~\ref{claim 3} together with item~\ref{cor2 - tk} in Corollary~\ref{corollary: stability}. Next, since $\tilde{V}(t) := \tfrac{1}{2}\tilde{e}_q^\top(t)\tilde{e}_q(t)$ either decreases exponentially or remains constant due to (\ref{eq: eq tilde and eq relation - general}) and item~\ref{cor2 - stability} in Corollary~\ref{corollary: stability}, item~\ref{thm4 - feasibility} holds by Corollary~\ref{corollary - feasibility}. Finally, combining (\ref{eq: eq tilde and eq relation - general}) with item~\ref{cor2 - stability} of Corollary~\ref{corollary: stability}, and taking $t \to \infty$, establishes item~\ref{thm4 - stability}.
\end{proof}

\begin{remark} \label{remark 5}
    In cooperative nonprehensile manipulation, when an event is triggered, the robots temporarily apply zero force to the object while moving to the newly switched input channels determined by $\delta$. Let $\Delta_k$ denote the time required for all robots to reach the new channels after the $k$-th event trigger. Then, under $\dot{q}_d = 0$, Theorem~\ref{theorem - with delay} guarantees stability of the closed-loop system even in the presence of such noninstantaneous switching.
\end{remark}
}

\section{Simulation Results} \label{section-simulation results}


\subsection{Validation of the proposed control algorithm} \label{subsection-results without robot}

\begin{figure}
    \centering
    \begin{subfigure}{0.45\linewidth}
        \centering
        \includegraphics[width=1.0\linewidth]{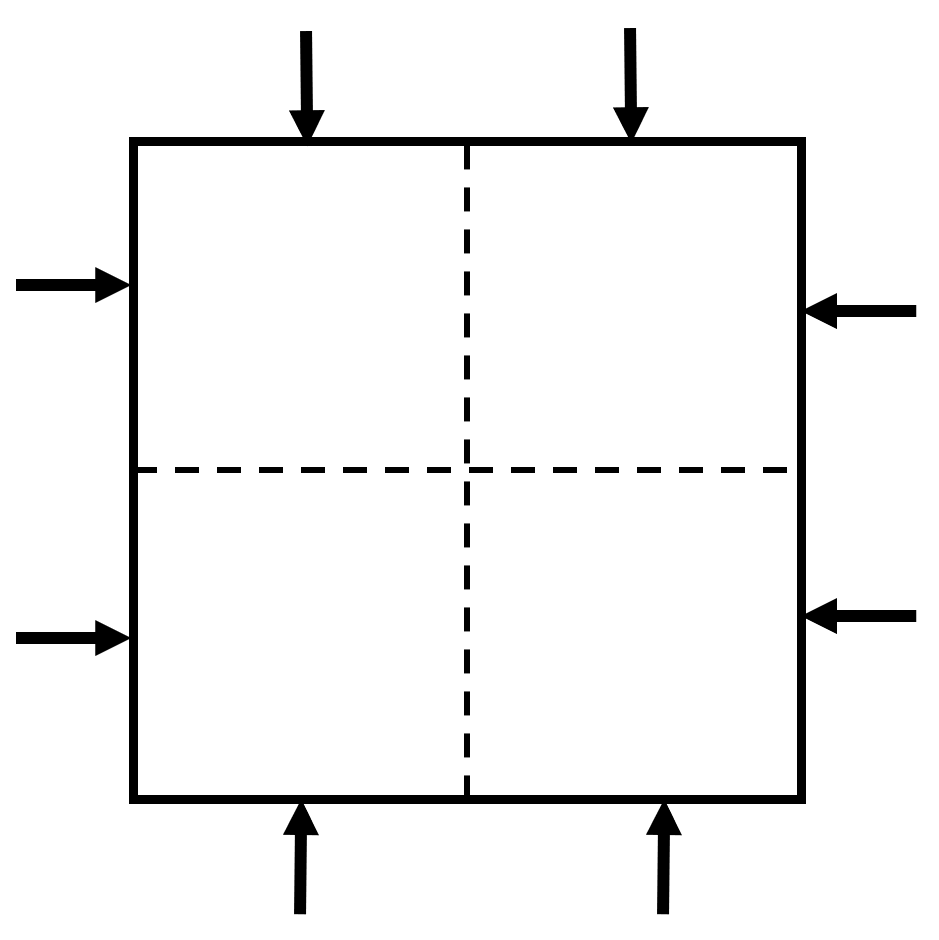}
        \caption{Square object}
        \label{fig:square object}
    \end{subfigure}
    \begin{subfigure}{0.45\linewidth}
        \centering
        \includegraphics[width=0.9\linewidth]{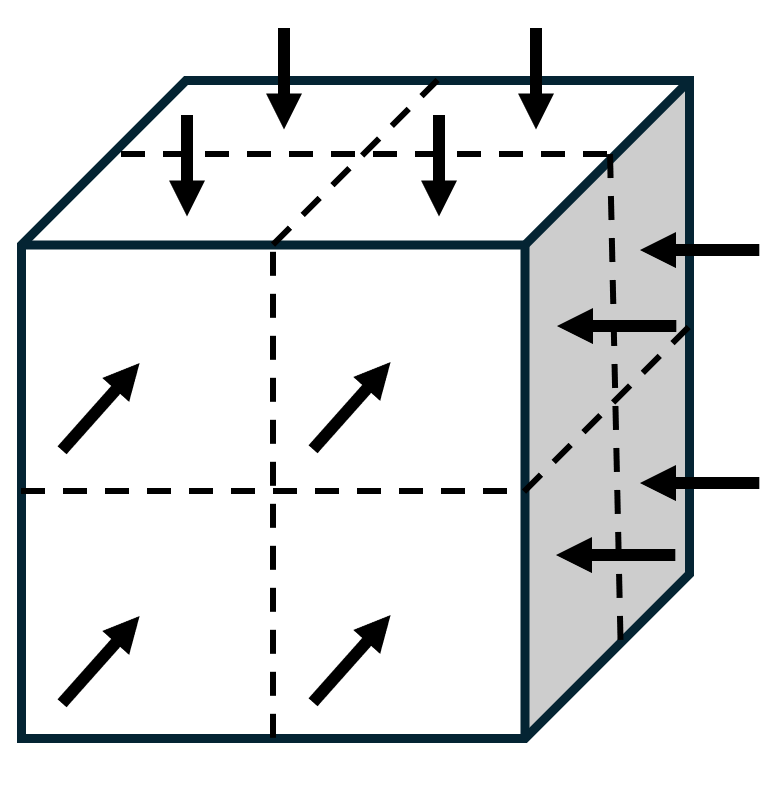}
        \caption{Cubic object}
        \label{fig:cubic object}
    \end{subfigure}
    \caption{Input channel configuration of examples used in simulation (left: square object with $8$ input channels, right: cubic object with $24$ input channels). The rest $12$ input channels not displayed in the cubic object are defined in a symmetric manner.}
    \label{fig:simulation setting}
\end{figure}

\begin{figure}
    \centering
    \includegraphics[width=0.8\linewidth]{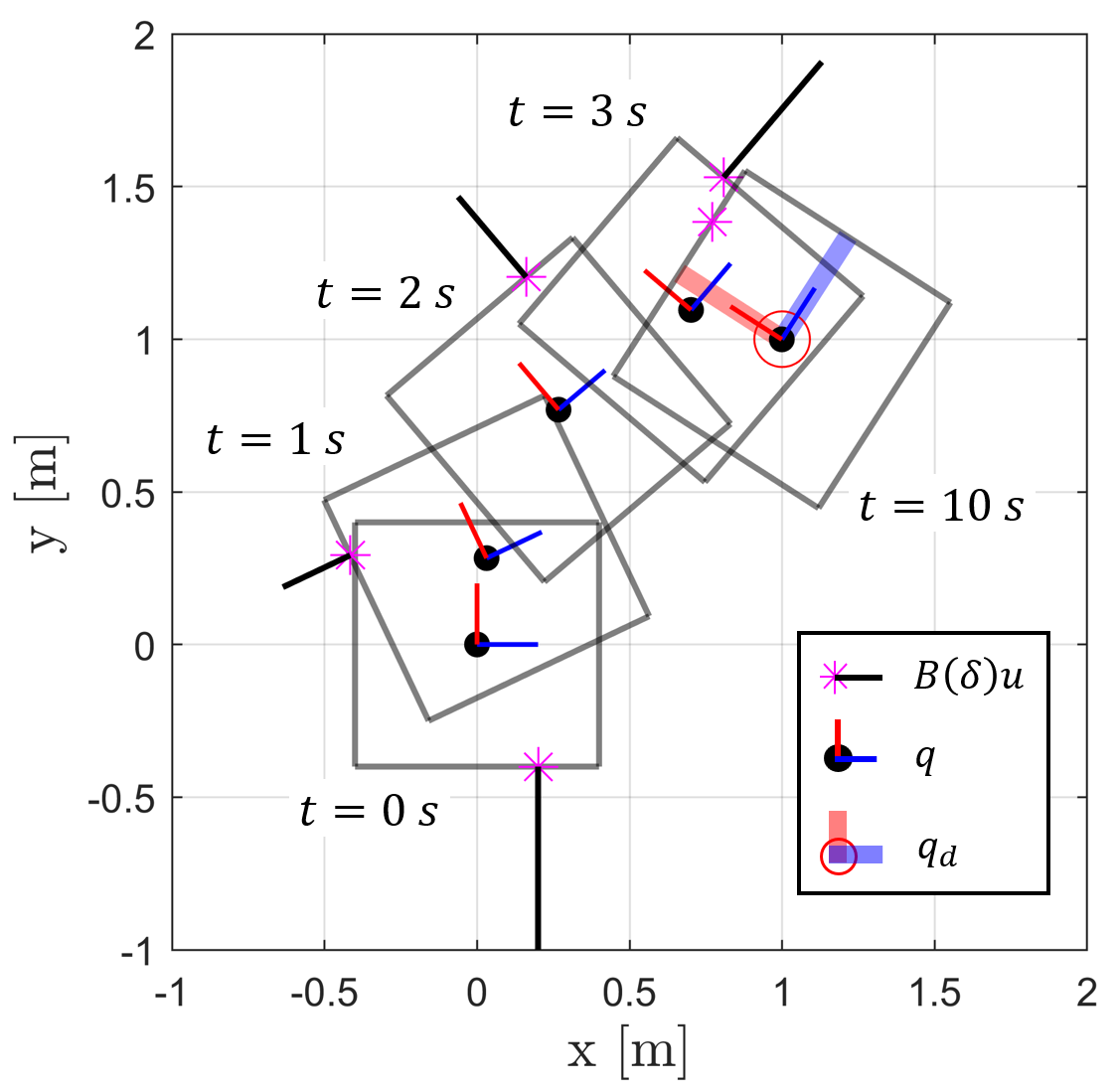}
    \caption{A composite image of simulation for Example 1 (square object) with $n_a = 1$.}
    \label{fig:simul_2d}
\end{figure}

We validate the proposed event-triggered switching controller in numerical simulation. Simulation codes are implemented in MATLAB and tested on a laptop computer equipped with Intel Core Ultra 5 125H CPU at 3.60 GHz and 32 GB RAM. Motivated by \cite[Ch. 10]{pedro2024predictive} and \cite{daley2020astrobee}, we select two simulation examples where the selected multi-channel systems emulating free-flyers with no gravity in 2D and 3D spaces are visualized in Fig. \ref{fig:simulation setting}. The positive span property (Definition \ref{def-positive span}) for the selected configurations of the input channels can be obtained by referring to \cite{regis2016properties}. The MILP in (\ref{eq: switching controller - MILP}) is solved using CPLEX \cite{cplex} whose average computation time is 0.08 \si{s} for both examples. Every simulation runs for $10$ \si{s}, and we set the discrete time interval to be $dt = 1$ \si{ms}.

\begin{figure}
    \centering
    \includegraphics[width=0.8\linewidth]{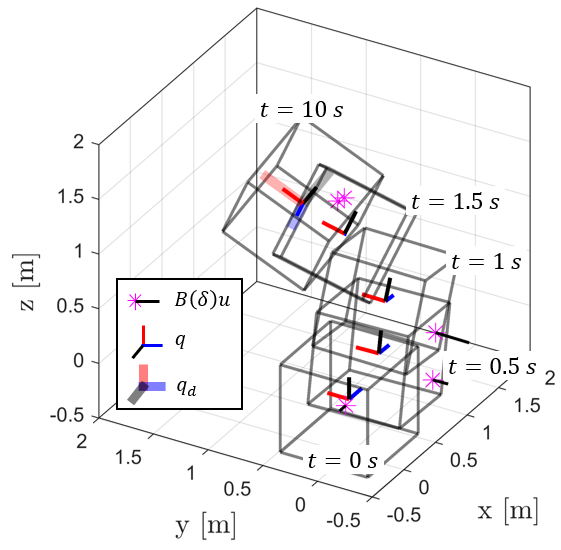}
    \caption{A composite image of simulation for Example 2 (cubic object) with $n_a = 1$.}
    \label{fig:simul_3d}
\end{figure}

In Figs. \ref{fig:square object simul result}, \ref{fig:cube object simul result}, to visualize which input channel is activated, we plot $\delta_{num}(t) = \texttt{diag}(\delta (t)) v_{n_a}$ where $v_{n} = [1;2;\cdots;n] \in \mathbb{R}^n$. Error variables $e_q, e_v$ and the Lyapunov function $V$ are illustrated, and $V_d(t)$, which is the solution of $\dot{V}_d = -c_d V_d$ with $V_d(0) = V|_{t=0}$ is also visualized along with $V$. Lastly, $\rho$ is the auxiliary optimization variable in MILP (\ref{eq: switching controller - MILP}) which is denoted in a logarithmic scale. Figs. \ref{fig:simul_2d} and \ref{fig:simul_3d} illustrate composite images of simulations at multiple timings. \blue{Videos of these simulation results can be found in \cite{youtube_video}.}

\subsubsection{Example 1 -- square object ($n=3$, $m=8$)} \label{subsec: example 1}

\begin{figure}
    \centering
    \includegraphics[width=1.0\linewidth, trim={0.8cm 0 0.8cm 0},clip]{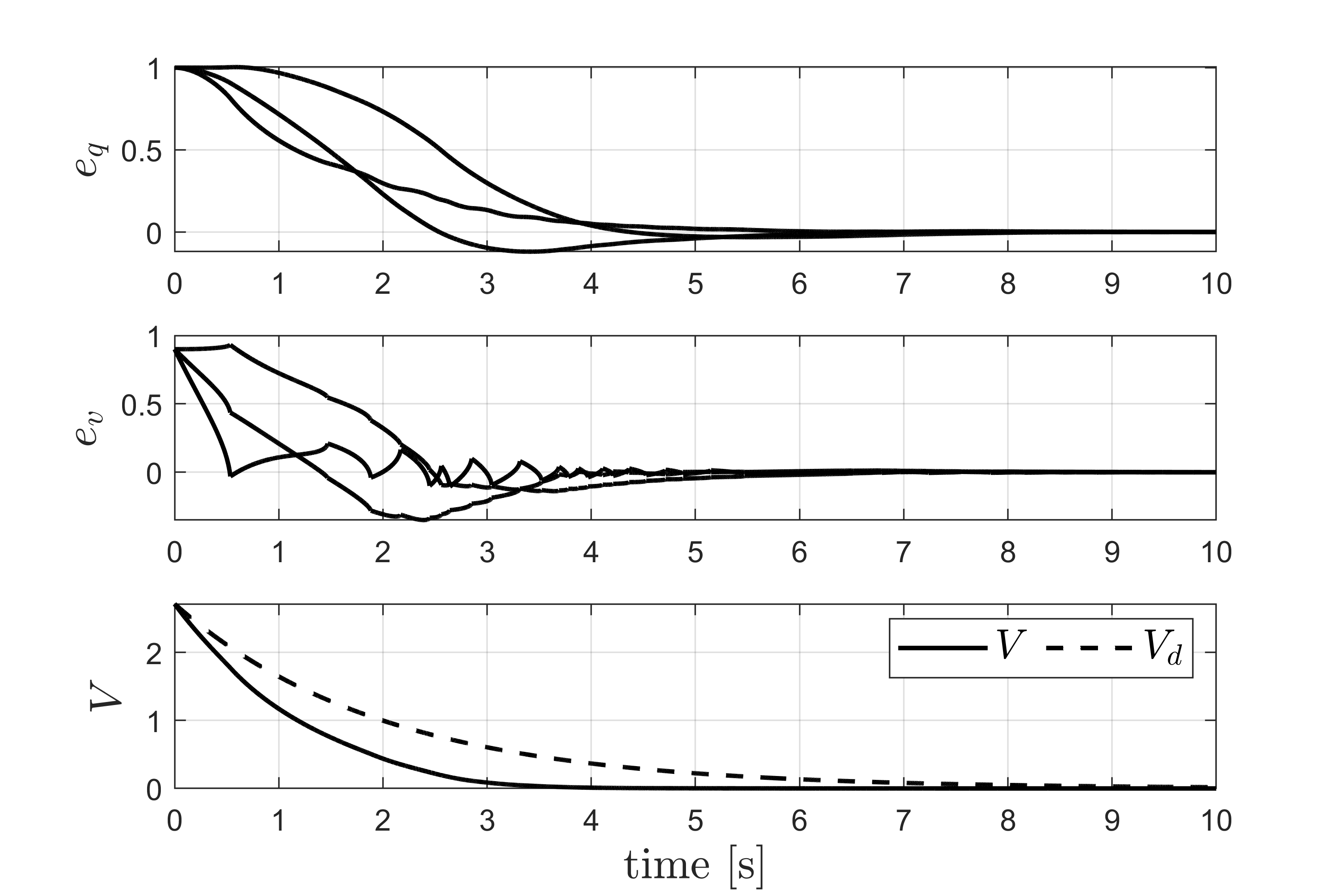}
    \includegraphics[width=1.0\linewidth, trim={0.8cm 0 0.8cm 0},clip]{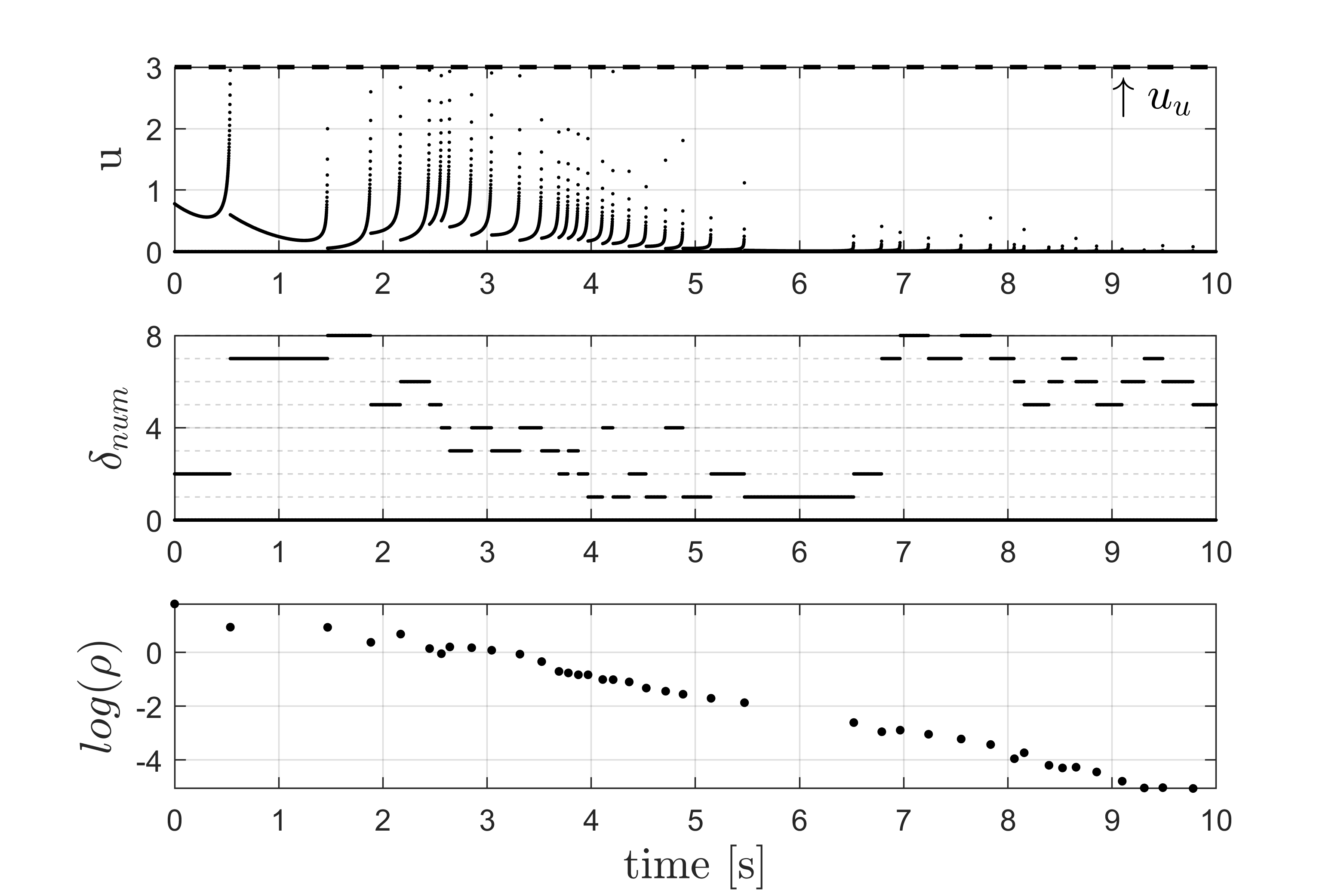}
    \caption{Example 1 -- square object with $n_a = 1$ and $u_u = 3 \ N$. The top two rows show the time evolution of the state errors, while the third row presents the Lyapunov function which always remains smaller than $V_d(t) = e^{-c_d t} V_0$, as analyzed in Theorem \ref{theorem: stability}. The fourth and fifth rows illustrate the control inputs, where $\delta_{num}(t) = \texttt{diag}(\delta (t)) v_{n_a}$ for $v_{n} = [1;2;\cdots;n] \in \mathbb{R}^n$. Finally, the bottom row displays the auxiliary variable $\rho$ in (\ref{eq: switching controller}) on a logarithmic scale, showing $\rho > 0$ if $V > 0$, as analyzed in Theorem \ref{theorem - feasibility}.}
    \label{fig:square object simul result}
\end{figure}

We consider a multi-channel system configured as Fig. \ref{fig:square object} where $q = [p_x;p_y;\theta] \in \mathbb{R}^3$ and $u \in \mathbb{R}^8$. The system dynamics can be written as
\begin{equation} \label{eq: square object dynamics}
    \ddot{q} = M^{-1} \begin{bmatrix}
        R & 0_2 \\ 0^\top_2 & 1
    \end{bmatrix} B(\delta) u,
\end{equation}
where $B(\delta) = \begin{bmatrix} b_1 \delta_1 & \cdots & b_8 \delta_8 \end{bmatrix} \in \mathbb{R}^{3 \times 8}$ is defined with $b_i$'s which are
\begin{equation} \label{eq: example 1 - bis}
\begin{gathered}
    b_1 = \begin{bmatrix}
        0 \\ 1 \\ -l
    \end{bmatrix}, b_2 = \begin{bmatrix}
        0 \\ 1 \\ l
    \end{bmatrix}, b_3 = \begin{bmatrix}
        -1 \\ 0 \\ -l
    \end{bmatrix}, b_4 = \begin{bmatrix}
        -1 \\ 0 \\ l
    \end{bmatrix}, \\
    b_5 = \begin{bmatrix}
        0 \\ -1 \\ -l
    \end{bmatrix}, b_6 = \begin{bmatrix}
        0 \\ -1 \\ l
    \end{bmatrix}, b_7 = \begin{bmatrix}
        1 \\ 0 \\ -l
    \end{bmatrix}, b_8 = \begin{bmatrix}
        1 \\ 0 \\ l
    \end{bmatrix}.
\end{gathered}
\end{equation}
Here, $l > 0$ is a constant length denoting the moment arm of each input channel. The mass matrix $M \in \mathbb{R}^{3 \times 3}$ and the rotation matrix $R \in \mathsf{SO}(2)$ are defined as
\begin{equation*}
    M = \begin{bmatrix}
        m I_2 & 0_2 \\ 0_2^\top & J
    \end{bmatrix}, \quad
    R = \begin{bmatrix}
        \cos(\theta) & -\sin(\theta) \\
        \sin(\theta) & \cos(\theta)
    \end{bmatrix},
\end{equation*}
where $m,J >0$ are mass and mass moment of inertia, respectively. Such dynamical model (\ref{eq: square object dynamics}) describes the motion of a planar free-flyer with multiple input channels \cite[Ch. 10]{pedro2024predictive}.

In simulation, the most extreme case, i.e. when $n_a = 1$, is tested with $u_u = 3$ \si{N}. Initial condition is given as $q = 0_3, v= 0_3$, and the desired state is defined as $q_d = 1_3, \dot{q}_d = \ddot{q}_d = 0_3$. As analyzed in (\ref{eq: V(t) exponential decrease}), $V(t) \leq V_d(t) = e^{-c_d t} V_0$ can be observed in the third row of Fig. \ref{fig:square object simul result}. Furthermore, the last row in Fig. \ref{fig:square object simul result} confirms that $\log(\rho) > -\infty$ exists, i.e. $\rho > 0$ as proved in Theorem \ref{theorem - feasibility}, whenever the switching controller (\ref{eq: switching controller - MILP}) is triggered. 



\subsubsection{Example 2 -- cubic object ($n=6$, $m=24$)} \label{subsec: example 2}

\begin{figure}
    \centering
    \includegraphics[width=1.0\linewidth, trim={0.8cm 0 0.8cm 0},clip]{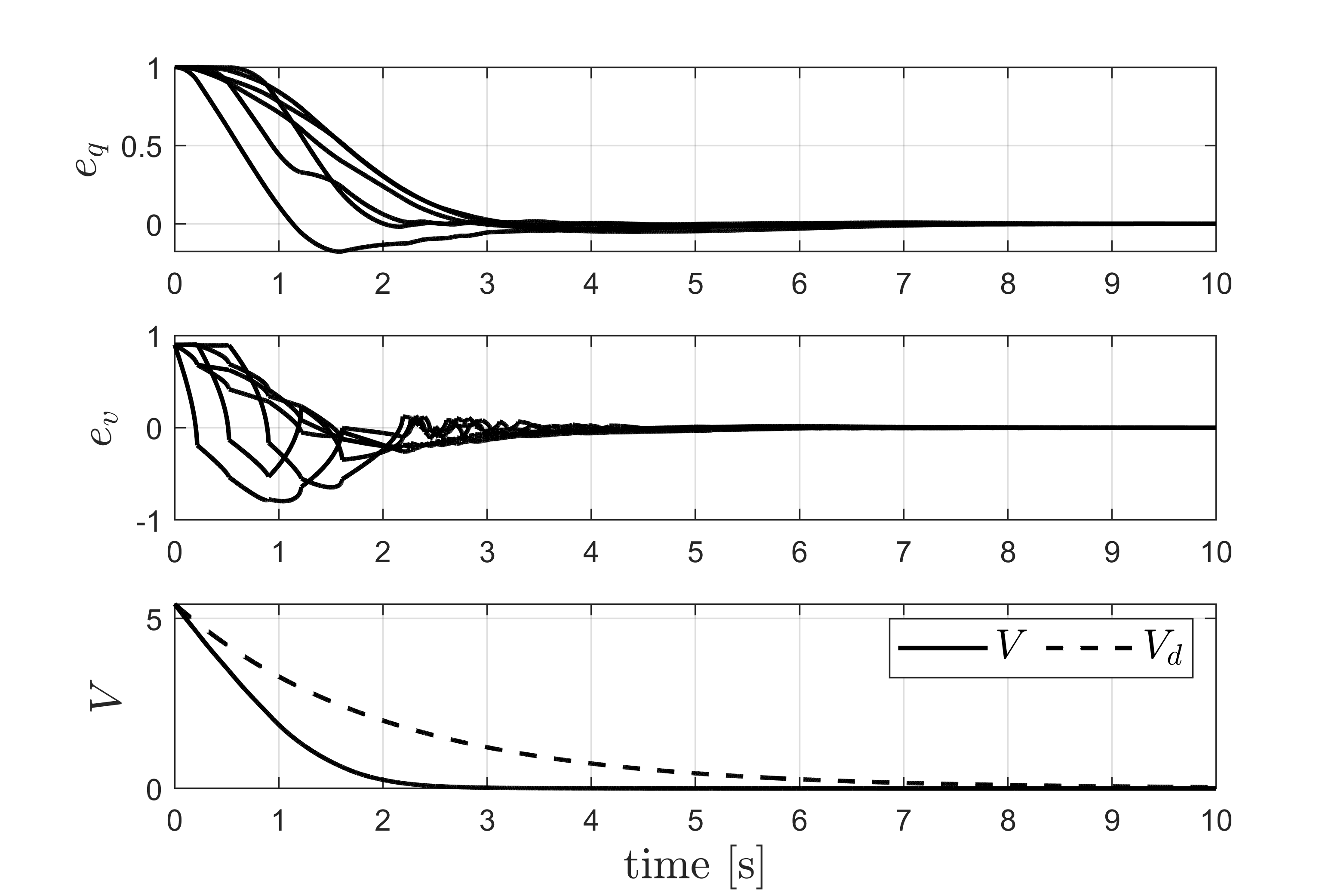}
    \includegraphics[width=1.0\linewidth, trim={0.8cm 0 0.8cm 0},clip]{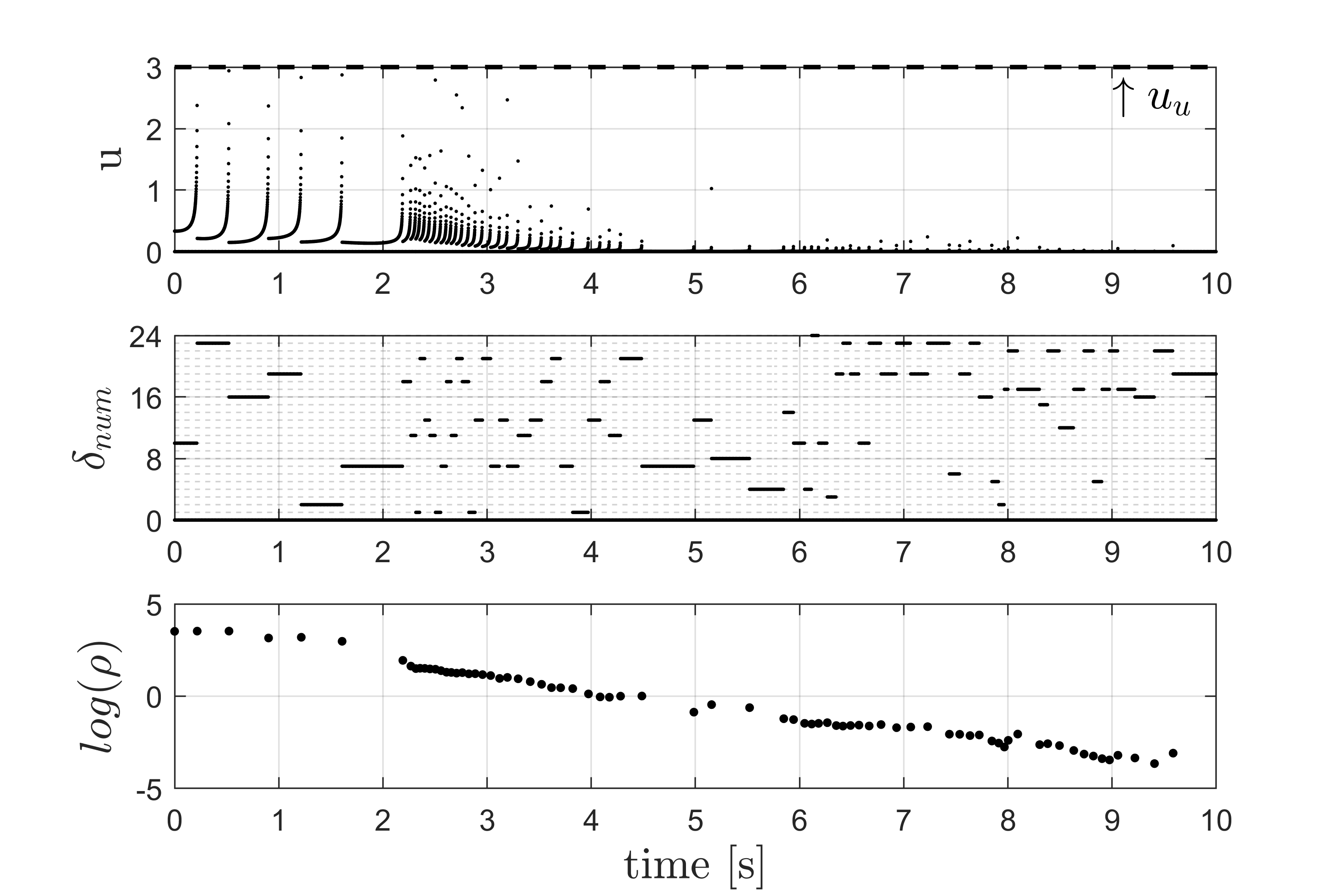}
    \caption{Example 2 -- cubic object with $n_a = 1$ and $u_u = 3 \ N$. The top two rows show the time evolution of the state errors, while the third row presents the Lyapunov function which always remains smaller than $V_d(t) = e^{-c_d t} V_0$, as analyzed in Theorem \ref{theorem: stability}. The fourth and fifth rows illustrate the control inputs, where $\delta_{num}(t) = \texttt{diag}(\delta (t)) v_{n_a}$ for $v_{n} = [1;2;\cdots;n] \in \mathbb{R}^n$. Finally, the bottom row displays the auxiliary variable $\rho$ in (\ref{eq: switching controller}) on a logarithmic scale, showing $\rho > 0$ if $V > 0$, as analyzed in Theorem \ref{theorem - feasibility}.}
    \label{fig:cube object simul result}
\end{figure}

To describe the free-floating cubic object in $3$-dimensional space, we define the configuration as $q = [p_x;p_y;p_z;\eta] \in \mathbb{R}^6$ where $\eta = [\phi;\theta;\psi] \in \mathbb{R}^3$ is ZYX Euler angles. Then, the system dynamics can be derived as follows:
\begin{equation*}
    \begin{bmatrix}
        m I_3 & 0_{3 \times 3} \\ 0_{3 \times 3} & J
    \end{bmatrix} \ddot{q} = \begin{bmatrix}
        0_3 \\ -(Q \dot{\eta})^\wedge J (Q \dot{\eta})
    \end{bmatrix} + \begin{bmatrix} R & 0_{3\times 3} \\ 0_{3\times 3} & I_3 \end{bmatrix} B(\delta) u,
\end{equation*}
where $(\cdot)^\wedge$ maps a vector in $\mathbb{R}^3$ to a skew-symmetric matrix in $\mathbb{R}^{3\times 3}$, $Q \in \mathbb{R}^{3 \times 3}$ is a Jacobian matrix between a body angular velocity $\omega \in \mathbb{R}^3$ and Euler angle rate $\dot{\eta}$ as $\omega = Q \dot{\eta}$. $R \in \mathsf{SO}(3)$ is the rotation matrix, and $m \in \mathbb{R}$ and $J \in \mathbb{R}^{3 \times 3}$ are mass and mass moment of inertia, respectively. Finally, $B(\delta) = [b_1 \delta_1 \ \cdots \ b_{24} \delta_{24}] \in \mathbb{R}^{6 \times 24}$. The constant vectors $b_i \in \mathbb{R}^6$ $\forall i = 1,\cdots,24$ can be obtained from geometric configuration of the input channels described in Fig. \ref{fig:cubic object}.

In simulation, we test the most extreme case, i.e. when $n_a = 1$, with $u_u = 3$ \si{N}. Similarly to Example 1, we set the initial condition as $q=0, v=0$, and the desired state is defined as $q_d = 1_6, \dot{q}_d = 0_6, \ddot{q}_d = 0_6$. $V(t) \leq V_d(t) = e^{-c_d t} V_0$ can be observed in the thrid row of Fig. \ref{fig:cube object simul result} as analyzed in (\ref{eq: V(t) exponential decrease}), and $\rho$ from the switching controller (\ref{eq: switching controller - MILP}) is always kept positive when the switching controller is triggered, which can be found in the bottom row of Fig. \ref{fig:cube object simul result}.

\subsubsection{Empirical analysis for different number of active agents}

In Table \ref{tab:number of events by na changes}, we empirically evaluate how the number of agents affects the number of event triggers. The number of event decreases as the number of active agents increases, and both the minimum time difference and the average time difference between the consecutive event triggers follow the similar tendency. Furthermore, considering that the minimal time differences between the consecutive event triggers of all data are larger than the discrete time interval of $dt = 1$ \si{ms}, we could see that event triggers does not occur consecutively in time. By comparing the total number of time discretization steps ($10000$) with the number of event triggers during the simulation, it is clear that the event-triggered control scheme in (\ref{eq: event-triggering condition}) and (\ref{eq: qp formulation}) is effective in reducing the number of MILPs (\ref{eq: switching controller - MILP}) to be solved, thereby improving the computation time.

\begin{table}
    \caption{Minimum and average time difference between consecutive event triggers and number of events for different number of agents.}
    \label{tab:number of events by na changes}
    \centering
    \begin{tabular}{@{}c c | c c c c c@{}} 
    \toprule
         \multicolumn{2}{c|}{$n_a$} & 1 & 2 & 4 & 8 & 16 \\
         \midrule
         \midrule
         \multirow{3}{*}{\shortstack{Example 1\\(square)}} & min [\si{ms}]     & 81 & 570 & 567 & - & - \\
                                             & average [\si{ms}] & 250.7 & 871.1 & 1669.7 & - & - \\
                                             & \# of events & 40 & 11 & 5 & 0 & - \\
         \midrule
         \multirow{3}{*}{\shortstack{Example 2\\(cube)}}   & min [\si{ms}]     & 37  & 43  & 74 & 125 & 1062 \\
                                             & average [\si{ms}] & 131.3 & 146.7 & 248.8 & 897.7 & 2135 \\
                                             & \# of events & 74 & 64 & 41 & 12 & 5 \\
    \bottomrule
    \end{tabular}    
\end{table}

\begin{figure*}[t]
    \centering
    \begin{subfigure}{\linewidth}
        \centering
        \includegraphics[width=\linewidth]{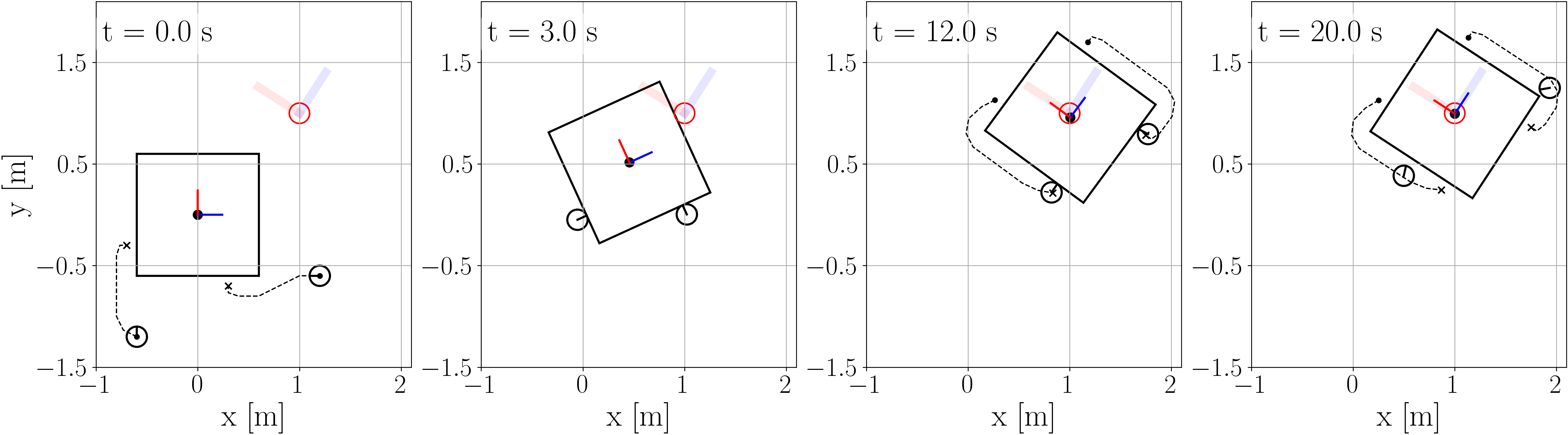}
        \caption{\blue{Nonprehensile cooperative manipulation using two robots.}}
        \label{fig:two_robots}
    \end{subfigure}
    \begin{subfigure}{\linewidth}
        \centering
        \vspace{2mm}
        \includegraphics[width=\linewidth]{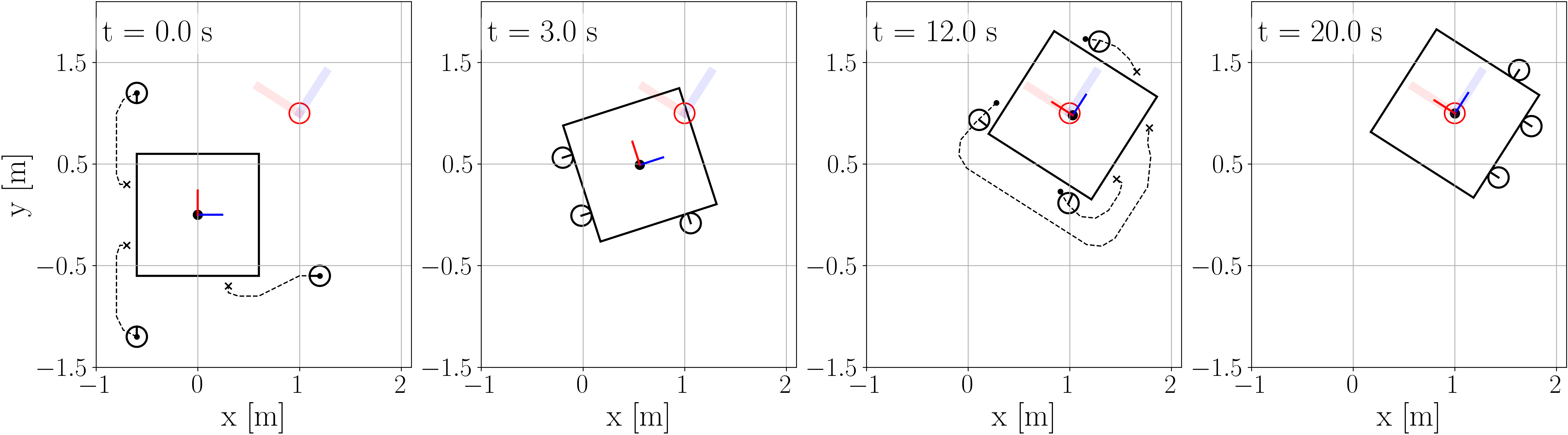}
        \caption{\blue{Nonprehensile cooperative manipulation using three robots.}}
        \label{fig:three_robots}
    \end{subfigure}

    \caption{\blue{Results of the nonprehensile cooperative manipulation in two different scenarios. The black circles represent the robots, and the black square denotes the object being manipulated. The red circle and the surrounding semi-transparent blue and red lines indicate the desired pose of the object. The black dashed lines shown at $t=0,12,20$~\si{s} represent the collision-free paths along which each robot navigates toward its newly assigned input channel.}}
    \label{fig:simulation_comparison}
\end{figure*}

{\color{blue}
\subsection{Application to nonprehensile cooperative manipulation}

We validated the proposed nonprehensile cooperative manipulation method using a \textit{square-shaped object}, as presented in Example~1 of the previous subsection. Note that the presented theorems hold for \textit{any object shape}, as long as Assumption~\ref{assumption - positive span} is satisfied by considering a sufficiently large number of input channels or contact points. The input channels used in this experiment are identical to those in Example~1 of the previous subsection, where $B(\delta) = [b_1 \delta_1 \cdots b_8 \delta_8] \in \mathbb{R}^{3 \times 8}$ and each $b_i$ is defined as in (\ref{eq: example 1 - bis}). Similar to the previous simulation, the initial and desired states of the manipulated object were set to $q = 0_3$ and $q_d = [1,\,1,\,1]$, respectively. The mass of the object was set to $10~\si{kg}$, and the static friction coefficient was $0.1$. The maximum pushing force of each robot was set to $u_u = 10~\si{N}$, corresponding to a small $1~\si{kg}$ robot capable of exerting a force approximately equal to its own weight. These parameters satisfy condition~(\ref{eq: upper bound condition - constant - qs}) required by Theorem~\ref{theorem - with delay}.

\begin{figure*}[t]
    \centering
    \begin{subfigure}{0.49\linewidth}
        \centering
        \includegraphics[width=\linewidth]{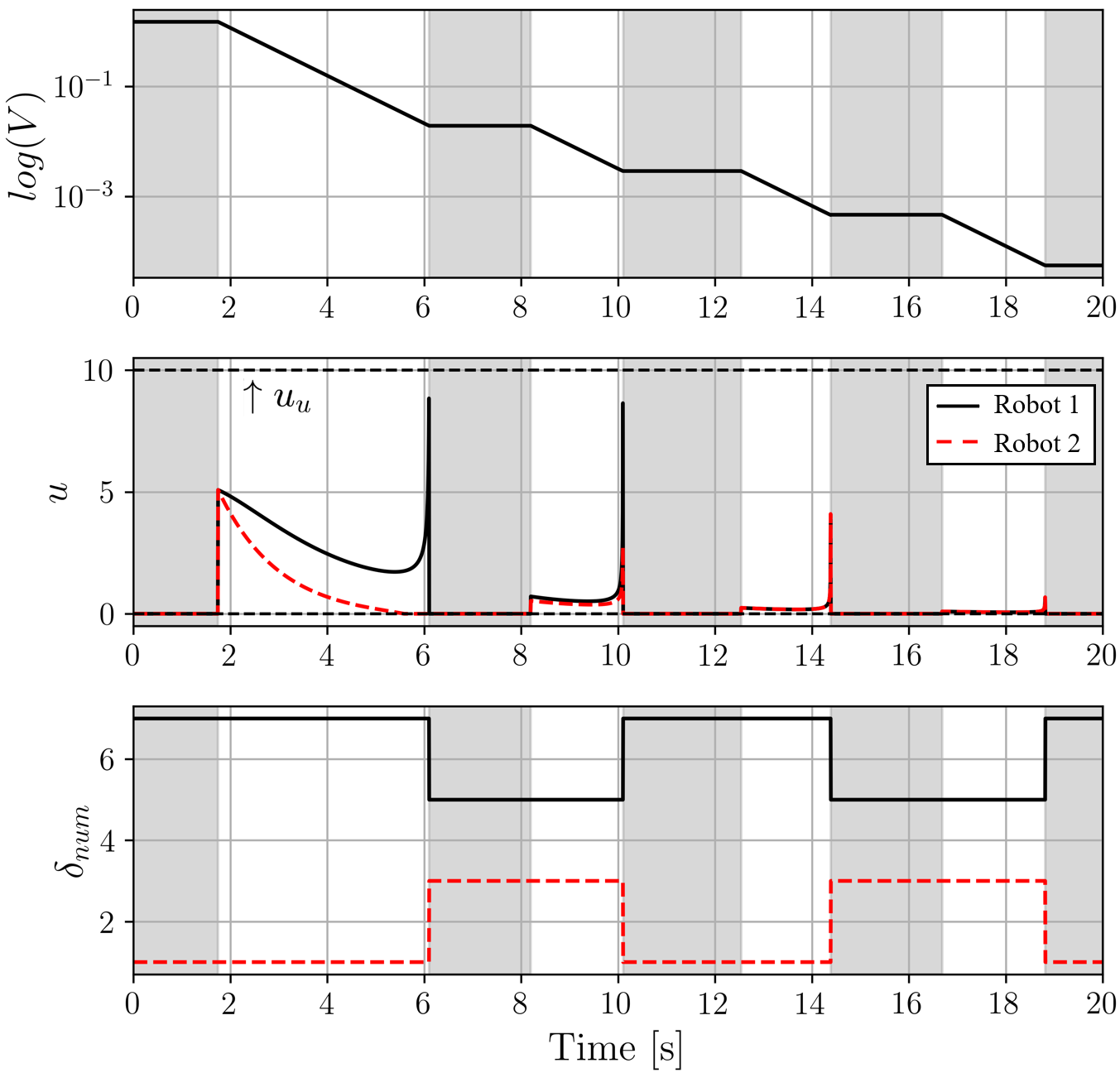}
        \caption{\blue{Nonprehensile cooperative manipulation using two robots.}}
        \label{fig:simulation with 2 robots}
    \end{subfigure}
    \begin{subfigure}{0.49\linewidth}
        \centering
        \includegraphics[width=\linewidth]{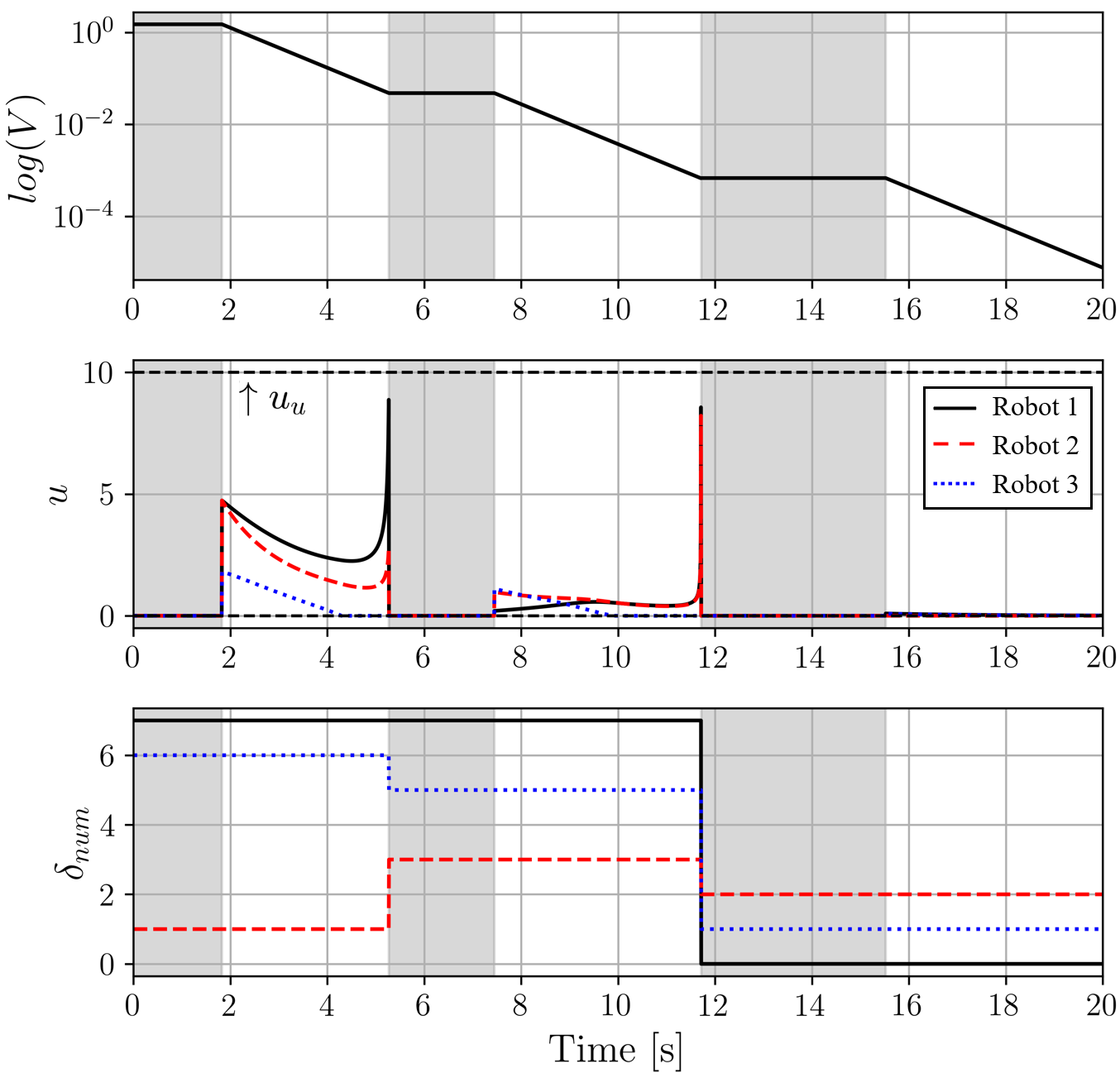}
        \caption{\blue{Nonprehensile cooperative manipulation using three robots.}}
        \label{fig:simulation with 3 robots}
    \end{subfigure}

    \caption{\blue{Results of the nonprehensile cooperative manipulation in two scenarios. The Lyapunov function $V$, the control inputs $u$ generated by each robot during manipulation, and the input channels $\delta_{num}$ for each robot are shown. The shaded gray regions indicate the periods when the robots were in the navigation mode, applying no force to the object.}}
    \label{fig:simulation_comparison_data}
\end{figure*}

Beyond the theoretical validation of the proposed controller presented in Subsection~\ref{subsection-results without robot}, we further apply the controller to a \textit{nonprehensile cooperative manipulation} task. In cooperative manipulation, the switching among multiple robots requires consideration of the \textit{noninstantaneous switching} effect, i.e., the non-negligible navigation time that occurs when each robot moves to a newly assigned input channel according to the switching signal. This aspect has already been addressed in Theorem~\ref{theorem - with delay} and Remark~\ref{remark 5}. Furthermore, when switching occurs, an additional challenge arises regarding how to assign each robot to an input channel (the allocation problem) and how to generate collision-free trajectories for the robots as they move toward their assigned channels (the multi-agent path-finding problem). These allocation and multi-agent path-finding problems are beyond the scope of this study. For the purpose of simulation, however, we employed two classical yet effective algorithms—namely, the \textit{Hungarian method}~\cite{kuhn1955hungarian} for allocation and the \textit{Hierarchical Cooperative A* algorithm}~\cite{silver2005cooperative} for path finding. Videos of these simulation results can be found in \cite{youtube_video}.

The simulation was performed on the same laptop as in Subsection~\ref{subsection-results without robot}. The MILP problem was solved using the CPLEX solver, and the total simulation time was $20$~\si{s} with a discrete time step of $dt = 5$~\si{ms}. Figs.~\ref{fig:two_robots} and~\ref{fig:three_robots} illustrate representative snapshots of the results obtained using two and three robots, respectively. The small black solid circles represent the robots, and the central black square denotes the object being manipulated. The desired state  of the object is indicated by a red solid circle for position and by a coordinate frame consisting of semi-transparent blue and red lines for orientation (corresponding to the $x$- and $y$-axes, respectively). In Fig.~\ref{fig:two_robots}, the black dashed lines with endpoints marked by \texttt{x} at $t=0,12,20$~\si{s}, and in Fig.~\ref{fig:three_robots} at $t=0,12$~\si{s}, represent collision-free trajectories along which each robot navigates toward its newly assigned input channel; the \texttt{x} marker indicates the newly assigned channel. In both simulations, the robots successfully cooperated to manipulate the object to its desired position and orientation by approximately $t=20$~\si{s}.

The variations of the control inputs, input channels, and Lyapunov function over the entire simulation are shown in Figs.~\ref{fig:simulation with 2 robots} and~\ref{fig:simulation with 3 robots}. The shaded gray regions indicate the intervals during which the robots were in the navigation mode, moving toward their newly allocated input channels while exerting no force on the object. The Lyapunov function $V$ is plotted on a logarithmic scale for better visibility. As shown, $V$ remains constant during navigation and decreases exponentially during manipulation, which appears as a linear decrease in the log scale, consistent with the proof in Theorem~\ref{theorem - with delay}. In both figures, the control input $u$ stays within the prescribed bounds, between the upper bound $u_u$ and the lower bound $0$, at all times. During the $20$~\si{s} simulation, a total of five switchings occurred in the two-robot case and three in the three-robot case, with the average MILP computation time remaining below $0.01$~\si{s}. The information on input-channel switching can be found through the variable $\delta_{\text{num}}$.


}

\section{Conclusion} \label{section-conclusion}

In this study, we proposed an event-triggered switching control method for a class of nonlinear underactuated multi-channel systems. The proposed control approach accounts for constraints on input direction and upper bounds on input magnitude, while also ensuring feasibility and stability. Our method could be applied to address the underactuation issues caused by an insufficient number of agents or agent failures in cooperative manipulation. Additionally, with minor modifications, it can serve as a low-level controller that guarantees goal convergence in nonprehensile manipulation.

\blue{We have theoretically and numerically shown that, under the quasi-static assumption in the application to nonprehensile cooperative manipulation, the proposed method does not induce fast switching and can ensure stability even with noninstantaneous switching. However, when the quasi-static assumption does not hold, switching events occur too frequently, posing challenges for direct application to real robotic systems. To address this limitation,} a potential direction for future research is integrating our method with offline global planning techniques, such as \cite{graesdal2024towards} to reduce the number of required switching instances while still ensuring real-time adaptation and goal convergence. 
To handle model uncertainties such as estimation errors in the contact point, we consider leveraging adaptive control techniques \cite{slotine1987adaptive} in (\ref{eq: backstepping ctrller}). \blue{Lastly, extending the proposed framework toward distributed control could address communication constraints and scalability issues inherent in centralized approaches.}

\section*{Appendix}

\subsection{Corollary to Theorem \ref{theorem - feasibility}} \label{appendix: corollary}
\begin{corollary}[from Theorem~\ref{theorem - feasibility}] \label{theorem - feasibility when pulling allowed}
    Assume (\ref{eq: upper bound condition}), and instead of Assumption \ref{assumption - positive span}, assume that $\mathcal{C}(B)$ linearly spans $\mathbb{R}^n$. Then, for any $n_a \geq 1$, there exists a tuple of $(\delta,u,\rho)$ satisfying $-u_u \leq u \leq u_u$, (\ref{eq: ICS-nonZeno term}), (\ref{eq: ICS-stability const}) and (\ref{eq: ICS-agent number const}). Furthermore, if $V > 0$, (\ref{eq: ICS-nonZeno term}) holds with strict inequality.
\end{corollary}
\begin{proof}
    The only difference in proving this statement compared to Theorem \ref{theorem - feasibility} is that the definition of $d$ in Lemma \ref{lemma - d, d1} should be modified to 
    \begin{equation*}
        d = \min_{\lVert e \rVert = 1} \left( \max_{b_i \in \mathcal{C}(B)} | b_i^\top e | \right).
    \end{equation*}
    For this newly defined $d$, positiveness of $d$ can still be proven by linear independence property. (i.e. $\forall e \in \mathbb{R}^n \backslash \{ 0\}$, $\exists b_i \in \mathcal{C}(B(1_n))$ such that $b_i^\top e \neq 0$) The rest of the proof follows that of Theorem \ref{theorem - feasibility}.
\end{proof}

\subsection{Proof of Claim \ref{claim: argmax}}
\begin{proof}
    If $|J| > m - n_a$ where $|J|$ denotes cardinality of the set $J$, then the claim is trivially true. 
    
    Next, consider the case when $|J| \leq m - n_a$. We divide this case into $|J| < n_a$ and $|J| \geq n_a$. For the first instance, assume $|J| < n_a$ and let the solution of (\ref{eq: switching controller}) be $u^*, \delta^*, \rho^*$. We prove Claim \ref{claim: argmax} by contradiction. Assume that $\delta^*_j = 0$ $\forall j \in J$ satisfying (\ref{eq: bj}). For an index set $K \subset [m]\backslash J$ satisfying $|K| = |J|$ and $\delta^*_k = 1$ $\forall k \in K$, let $\delta^{new} \in \mathbb{Z}^m_2$, $u^{new} \in \mathbb{R}^m$ be defined as 
    \begin{equation*}
        \delta^{new}_i = \begin{cases}
            \delta^*_i & \text{if } i \notin (J \cup K) \\
            1 & \text{if } i \in J \\
            0 & \text{if } i \in K
        \end{cases},
        u^{new}_i = \begin{cases}
            u^*_i & \text{if } i \notin (J \cup K) \\
            u_u & \text{if } i \in J \\
            0 & \text{if } i \in K
        \end{cases}.
    \end{equation*}
    By definition, $u^{new}$ and $\delta^{new}$ satisfy (\ref{eq: ICS-pushing const}) and (\ref{eq: ICS-agent number const}). Next, since (\ref{eq: switching controller}) maximizes $\rho$ and there exists only one constraint imposing an upper bound on $\rho$, i.e. (\ref{eq: ICS-stability const}), the solution $\rho^*$ satisfies
    \begin{equation*}
        \rho^* = s_b^\top B(\delta^*) u^* -r,
    \end{equation*}
    by referring to (\ref{eq: stability const - modified}). Similarly, let $\rho^{new}$ be defined as $\rho^{new} = s_b^\top B(\delta^{new}) u^{new} -r$. Then, we can derive the following:
    \begin{equation} \label{eq: rho new, rho star inequality}
    \begin{gathered}
        \rho^{new} - \rho^* = s_b^\top \sum_{i \in [m]} b_i (\delta^{new}_i u^{new}_i -\delta^{*}_i u^{*}_i) \\
        = \sum_{j \in J} s_b^\top b_j u_u - \sum_{k \in K} s_b^\top b_k u^*_k \geq |J| (s_b^\top b_{j^*} u_u - \max_{k\in K} s_b^\top b_k u^*_k) \\
        \geq |J| (s_b^\top b_{j^*} u_u - \max \{0, \max_{k \in K} s_b^\top b_k u_u \}) > 0,
    \end{gathered}
    \end{equation}
    where $j^*$ is any element in $J$, and we use the fact that $u_u \geq 0$, $s_b^\top b_{j^*} > 0$ which can be obtained from Lemma \ref{lemma - d, d1} using $d_1 > 0$ and the definition of $J$ in (\ref{eq: bj}). The new solution $u^{new}, \delta^{new}, \rho^{new}$ satisfying all the constraints (\ref{eq: ICS-pushing const}), (\ref{eq: ICS-nonZeno term}), (\ref{eq: ICS-stability const}) and (\ref{eq: ICS-agent number const}) shows a lower value of the cost function in (\ref{eq: switching controller}) than the original solution $u^*, \delta^*, \rho^*$ as derived in (\ref{eq: rho new, rho star inequality}); thus, this contradicts to the fact that $u^*, \delta^*, \rho^*$ is the solution to (\ref{eq: switching controller}). By following a similar procedure, this contradiction can also be derived for the other instance where $|J| \geq n_a$, and thus, we omit the proof for it.
\end{proof}

\normalem


\end{document}